\documentclass[11pt]{article}
\pdfoutput=1
\usepackage{jheppub}
\usepackage{physics}
\usepackage{bbm}
\usepackage{comment}

\preprint{YITP-SB-2023-24}

\newcommand\nn{\nonumber}

\newcommand{\wt}{\widetilde}

\usepackage{tikz}
\usetikzlibrary{arrows,snakes,shapes.arrows,decorations.markings}
     \tikzset{>=triangle 90}
     \tikzstyle{bbc}=[draw,circle,fill=black,scale=.75]
     \tikzstyle{rc}=[circle,fill=red,scale=.6]
     \tikzstyle{wc}=[draw,circle,scale=.75]

\def\bar{\overline}

\def\^{\wedge}

\def\g{{\gamma}}

\def\G{{\Gamma}}

\def\t{{\tau}}

\def\cD{{\mathcal D}}

\def\cH{{\mathcal H}}

\def\cI{{\mathcal I}}

\def\cN{{\mathcal N}}

\def\cS{{\mathcal S}}

\def\cT{{\mathcal T}}

\def\cV{{\mathcal V}}

\def\E{\mathbb{E}} 
\def\C{\mathbb{C}} 
 
\def\H{\mathbb{H}}
\def\N{\mathbb{N}}

\def\R{\mathbb{R}} 
 
\def\Z{\mathbb{Z}} 

\def\beq{\begin{equation}}
\def\eeq{\end{equation}}
\def\nn{\nonumber}

\begin{document}

\title{High-temperature expansion of the Schur index and modularity}

\author[a]{Arash Arabi Ardehali,}
\affiliation[a]{C.N.~Yang Institute for Theoretical Physics, Stony Brook University,\\ Stony Brook, NY 11794, USA}
\author[b]{Mario Martone,}
\author[b]{Mart\'i Rossell\'o}
\affiliation[b]{Department of Mathematics, King's College London,\\
The Strand, London WC2R 2LS, U.K.}

\emailAdd{a.a.ardehali@gmail.com}\emailAdd{mario.martone@kcl.ac.uk}\emailAdd{marti.rossello@kcl.ac.uk}

\abstract{High-temperature ($q\to1$) asymptotics of 4d superconformal indices of Lagrangian theories have been recently analyzed up to exponentially suppressed corrections. Here we use RG-inspired tools to extend the analysis to the exponentially suppressed terms in the context of Schur indices of $\mathcal{N}=2$ SCFTs. In particular, our approach explains the curious patterns of logarithms (polynomials in $1/\log q$) found by Dedushenko and Fluder in their numerical study of the high-temperature expansion of rank-$1$ theories.  We also demonstrate compatibility of our results with the conjecture of Beem and Rastelli that Schur indices satisfy finite-order, possibly twisted, modular linear differential equations (MLDEs), and discuss the interplay between our approach and the MLDE approach to the high-temperature expansion. The expansions for $q$ near roots of unity are also treated. A byproduct of our analysis is a proof (for Lagrangian theories) of rationality of the conformal dimensions of all characters of the associated VOA, that mix with the Schur index under modular transformations.}

\maketitle \flushbottom

\section{Introduction}

Asymptotics of supersymmetric partition functions of superconformal field theories (SCFTs) in $d>2$ dimensions have been under intense study in recent years, in particular due to the revival of AdS/CFT black hole microstate counting in \cite{Benini:2015eyy}. Supersymmetric partition functions are protected against quantum corrections and thus serve as an ideal means of exploring weak/strong dualities such as AdS/CFT, where their asymptotics in deconfined regimes allow making contact with bulk black holes \cite{Witten:1998zw}.  See \cite{Zaffaroni:2019dhb} for a review of the related developments until 2019.

The best understood class of such partition functions are superconformal indices \cite{Kinney:2005ej,Romelsberger:2005eg} 
\begin{equation}
\begin{split}
    &\mathcal{I}^{}_{\mathcal{N}=1}(p,q),\\
    &\mathcal{I}^{}_{\mathcal{N}=2}(p,q;v),\\
    &\mathcal{I}^{}_{\mathcal{N}=4}(p,q;y_1,y_2),
\end{split}\label{eq:4dIndices}
\end{equation}
of four dimensional SCFTs, and their best understood limit is the Cardy-like ($p,q\to1$, or ``high-temperature'') limit. A partial list of relevant references is \cite{Choi:2018hmj,Honda:2019cio,ArabiArdehali:2019tdm,Kim:2019yrz,Cabo-Bizet:2019osg,Amariti:2019mgp,ArabiArdehali:2019orz,GonzalezLezcano:2020yeb,Goldstein:2020yvj,Amariti:2020jyx,Lezcano:2021qbj,Amariti:2021ubd,Cassani:2021fyv,Ardehali:2021irq,Ohmori:2020wpk,DiPietro:2014bca,ArabiArdehali:2015iow,Buican:2015ina,ArabiArdehali:2015ybk,DiPietro:2016ond,Chen:2023lzq,Amariti:2023rci}.

A general analysis was presented in \cite{Ardehali:2021irq} which yields asymptotics of the indices of Lagrangian 4d SCFTs up to exponentially small corrections.\footnote{A similarly accurate result was proposed in \cite{Cassani:2021fyv} (improving earlier formulas in \cite{Cabo-Bizet:2019osg,Kim:2019yrz}) for the special index $\mathcal{I}^{}_{\mathcal{N}=1}(p,q\,e^{\pm2\pi i})$, informally referred to as the ``second-sheet'' index, which is valid when $p,q$ approach 1 in a certain range of angles such that the result corresponds to exponential growth (and not suppression).} In this work we extend that analysis to the exponentially suppressed terms in the special case of the index
\begin{equation}
    \mathcal{I}(q):=\mathcal{I}^{}_{\mathcal{N}=2}(q,q;q^{1/3})=\mathrm{Tr}(-1)^F\, q^{\Delta-R},
\end{equation}
known as the Schur index. In the above relation the trace is in the radial quantization, and $F,\Delta,R$ are the fermion number, conformal dimension, and $\mathrm{SU}(2)_{R_{\mathcal{N}=2}}$ charge respectively.

Even though the Schur index does not have a deconfined phase associated to bulk black holes (\emph{i.e.} it does not encode large black hole microstates, see \emph{e.g.} \cite{Eleftheriou:2022kkv}), it is of special interest because it coincides with the vacuum character of the 2d VOA associated to the 4d SCFT \cite{Beem:2013sza}. It also has analytic properties that are simpler than the more general indices in \eqref{eq:4dIndices}. In particular, exponential integrals that arise in the high-temperature expansion of more general indices may have piecewise quadratic functions in their exponent \cite{Ardehali:2021irq}, while for the Schur index the functions in the exponents are always piecewise linear; see Section~\ref{sec:tauTo0}. We will benefit greatly from this simplicity below in extending the analysis of \cite{Ardehali:2021irq} to the exponentially suppressed terms in the Schur index of SCFTs.\footnote{The Schur index has been studied also for non-conformal $\mathcal{N}=2$ theories (\emph{e.g.} in \cite{Cordova:2015nma}). In that context the functions in the exponents of the exponential integrals arising in the high-temperature expansion are no longer piecewise linear, and the simplicity discussed in the present paragraph is lost.}

Our starting point is the matrix integral expression for the Schur index---see Eq.~\eqref{eq:indexTheta0}---as an integral over a BPS moduli space. The integrand involves theta functions of $q=e^{2\pi i \tau}$ which, thanks to their modular properties, can be $S$-transformed into functions of $\tilde{q}:=e^{-2\pi i/\tau}$. The resulting functions can be expanded in $\tilde{q}$, yielding a high-temperature ($\tilde{q}\to0$) expansion for the integrand. However, the naive $\tilde{q}$ expansion is not uniformly convergent everywhere on the moduli space, and this can cause problem for the integral. As in \cite{Ardehali:2021irq} we confront this problem by decomposing the moduli space into an \emph{outer patch} where the naive expansion is uniformly valid, and various \emph{inner patches} where extra care is required.

Our outer patch analysis, which is a straightforward generalization of the outer patch analysis in \cite{Ardehali:2021irq}, already explains the curious patterns of logarithms encountered in the $\tilde{q}$ expansion of the Schur index by Dedushenko and Fluder in \cite{Dedushenko:2019yiw}. In our approach the logarithms arise from flat directions in various effective potentials (the piecewise linear functions in the exponents of the exponential integrals) over the BPS moduli space; specifically, the Rains function \cite{ArabiArdehali:2015ybk,Rains:2006dfy} and certain descendants of it that we will call \emph{higher Rains functions} in this work. See in particular the $\mathrm{SU}(2)$ $\mathcal{N}=4$ theory and $\mathrm{SU}(2)$ SQCD examples in Section~\ref{sec:tauTo0}.

The inner patches, on the other hand, were treated in \cite{Ardehali:2021irq} via subtraction methods that are not powerful enough on their own to give the full $\tilde{q}$ expansion that we want in this work. New tools are needed, and we find an algorithm inspired by renormalization group (RG) ideas that fits the bill; this is the main technical novelty of our work and is presented~in~Section~\ref{subsubsec:inner}.

Let us briefly illustrate this technical result with a simple toy model. Imagine we want to produce the small-$\tau$ (hence small-$\tilde{q}$) expansion of the following integral:
\begin{equation}
    \mathcal{I}(q)=\frac{2\pi i}{\tau}\int_0^1 \mathrm{d}x \frac{1}{(1+\tilde{q}^{x})(1+\tilde{q}^{1-x})}.\label{eq:toy}
\end{equation}
The range $[0,1]$ in the above integral is analogous to the integration domains of our Schur indices below, which are Cartan tori that are similarly compact.

The integral \eqref{eq:toy} itself can be evaluated in terms of hypergeometric functions, and then expanded around $\tilde{q}=0$ using the theory of hypergeometric functions. Here we want to avoid special function theory, as it is not available to the best of our knowledge for the integrals we will encounter studying higher-rank Schur indices (and in any case this part of our work is along a larger Wilsonian attempt to break away from confines of rigid analytics). Simply Taylor expanding the integrand can not work because the expansion is not allowed near the end-points $x=0,1$ where the powers of $\tilde{q}$ vanish and we approach the boundary of the domain of convergence of the series $1/(1+X)=1-X\pm\cdots .$ What we do instead is to decompose the integration domain into an outer patch $[\epsilon,1-\epsilon],$ and two inner patches $\text{in}_1=[0,\epsilon)$ and $\text{in}_2=(1-\epsilon,1]$. The numerical value of $\epsilon$ is chosen suitably small but fixed, for instance $\epsilon=0.2.$ The outer patch contribution is found easily, via Taylor expansion of the integrand. The inner patch contributions, on the other hand, are found via a combination of re-scaling and RG equations as follows. Re-scaling allows us to write:
\begin{equation}
\begin{split}
    \mathcal{I}_{\text{in}_1}(q)&=\int_0^\epsilon  \frac{\frac{2\pi i }{\tau}\mathrm{d}x}{(1+\tilde{q}^{x})(1+\tilde{q}^{1-x})}=\int_0^{\frac{2\pi i\epsilon}{\tau}}  \frac{\mathrm{d}x}{(1+e^{-x})(1+\tilde{q}e^{x})}=\sum_{n=0}^\infty (-\tilde{q})^n \, I_n\big(\frac{2\pi i\epsilon}{\tau}\big),    \end{split}\label{eq:toyInIntro}
\end{equation}
where the last equality follows from $1/(1+\tilde{q}e^x)=\sum_{n=0}^\infty (-\tilde{q}e^x)^n$, which is uniformly valid over in$_1$. The $I_n$ are defined as
\begin{equation}
I_n(\Lambda):=\int_0^{\Lambda}  \frac{e^{nx}\, \mathrm{d}x}{1+e^{-x}}.
\end{equation}
With all the non-trivial $\tau$-dependence encapsulated in the cut-off of $I_n(\Lambda)$ thanks to the re-scaling, an RG equation now allows us to get away from the problematic end $x=0$ of in$_1$:
\begin{equation}
    \frac{\mathrm{d}}{\mathrm{d}\Lambda}I_n(\Lambda)=\frac{e^{n\Lambda}}{1+e^{-\Lambda}}= e^{n\Lambda}\,\sum_{m=0}^\infty (-e^{-\Lambda})^m.\label{eq:I_nToy}
\end{equation}
Integrating the expansion on the RHS\footnote{Analogously to how integrating RGEs solves the problem of large logs in 4d QED \cite{Burgess:2020tbq} or $\varphi^4$ theory \cite{brezin2010introduction}.} with respect to $\Lambda$ gives $I_n(\Lambda)$, modulo only an integration constant which in turn can be found via the subtraction method---see footnote~\ref{ftnt:subtractionMethod}. Inserting that expansion into \eqref{eq:toyInIntro} gives $\mathcal{I}_{\text{in}_1}.$ The expansion of  $\mathcal{I}_{\text{in}_2}$ is found similarly. 

In other words, all the subtlety regarding the lack of uniform convergence near the end-points $x=0,1$ is encoded in the integration constants (analogous to Wilson coefficients), which in the present context can be determined via the subtraction method. Compiling $\mathcal{I}_{\text{in}_1},\mathcal{I}_{\text{in}_2},\mathcal{I}_{\text{out}}$ gives the full expansion of $\mathcal{I}(q)$. Section~\ref{subsubsec:inner} generalizes these ideas to higher~rank.

A complementary approach to the high-temperature expansion of Schur indices is via the (possibly twisted) modular linear differential equations (MLDEs) that they satisfy \cite{Arakawa:2016hkg,Beem:2017ooy} (see also \cite{Zheng:2022zkm}). More precisely, the Schur index $\cI(q)$ can be mapped to the vacuum character $\chi^{}_0(q)$ of the 2d VOA associated with the $\cN=2$ SCFT via
\begin{equation}
    \chi^{}_0(q):=q^{\mathrm{c}/2}\mathcal{I}(q).
\end{equation}
Here $\mathrm{c}$ is the four dimensional central charge. The 2d VOAs that arise from 4d SCFTs are in turn conjectured to be of a very special type known as ``quasi-lisse" \cite{Beem:2017ooy}, a property which ensures that their vacuum characters satisfy a---possibly twisted---MLDE \cite{Arakawa:2016hkg}. Now let us consider the case where the MLDE is untwisted (the twisted case is only marginally more complicated and will be discussed in Section~\ref{subsec:lmde}). Say we have an $n$th order MLDE
\begin{equation}
    \mathcal{D}^{}_q\, \chi^{}_0(q)=0.\label{eq:MLDE}
\end{equation}
Let $\chi_0,\dots,\chi_{n-1}$ be the $n$ linearly independent solutions to the above MLDE: $\mathcal{D}^{}_q\, \chi^{}_j(q)=0$. Changing variables they trivially also satisfy
\begin{equation}
    \mathcal{D}^{}_{\tilde{q}}\, \chi^{}_j(\tilde q)=0.\label{eq:tildedMLDE}
\end{equation}
But modularity implies that we can replace $\mathcal{D}^{}_{\tilde{q}}$ with $\mathcal{D}^{}_{{q}},$ so they moreover satisfy $\mathcal{D}^{}_{{q}}\, \chi^{}_j(\tilde q)=0$. Consequently \eqref{eq:MLDE} guarantees that $\chi^{}_0(q)$ is in their linear span: 
\begin{equation}
    \chi^{}_0(q)=\sum_{j=0}^{n-1} c_j \chi_j(\tilde{q}).\label{eq:ZinTermsOfMixedFj}
\end{equation}
We refer to the $\chi^{}_j$ with $c_j\neq0$ informally as the ones that ``mix with'' $\chi^{}_0$ under $\mathrm{SL}(2,\mathbb{Z}).$ Now the Fuchsian series solutions to \eqref{eq:tildedMLDE} of the form (assuming non-degeneracy mod $\mathbb{Z}$ of the roots $\alpha_j$ of the indicial equation)
\begin{equation}
    \chi^{}_j(\tilde{q})=\tilde{q}^{\alpha_j}\big(a_{j,0}+a_{j,1} \tilde{q}+\dots\big),
\end{equation}
yield the
$\tilde{q}$ expansion of $\chi^{}_0(q)$ via \eqref{eq:ZinTermsOfMixedFj}:
\begin{equation}
     \chi^{}_0(q)=\sum_{j=0}^{n-1} c_j \tilde{q}^{\alpha_j}\big(a_{j,0}+a_{j,1} \tilde{q}+\dots\big).
\end{equation}
In particular, the roots of the indicial equation of the MLDE \eqref{eq:MLDE} set the powers of $\tilde{q}$ appearing in the expansion, modulo positive integers. The smallest indicial root is hence expected to set the leading asymptotic in the high-temperature limit---supposing as we always do below that its corresponding $a_{j,0}$ and $c_j$ coefficients are nonzero.

Logarithmic terms (polynomials in $\log\tilde{q}$) in the high-temperature expansion arise from this perspective whenever there are repeated indicial roots, or roots that differ by integers. This way we get an alternative explanation for the patterns of logarithms encountered in~\cite{Dedushenko:2019yiw}. Note that since the MLDE implies $\chi_j(q)$ mix among themselves under modular transformations, the vector $\big(\chi_0(q),\dots,\chi_{n-1}(q)\big)^T$ is a vector-valued modular form (vvmf), so $\chi^{}_0(q)$ itself is a component of a vvmf. (More exactly a logarithmic vvmf when there are degenerate indicial roots mod $\mathbb{Z}$.)

One of the motivations for the present work was to investigate the interplay between the above two approaches to the $\tilde{q}$ expansion of Schur indices. Applying the two methods to various examples in Section~\ref{sec:tauTo0}, we find instances where one approach can efficiently constrain the results of the other. For example, in $\mathrm{SU}(2)$ SQCD the direct matrix integral approach initially suggests presence of certain powers of $\tilde{q}$ in the expansion that, without a more detailed examination one would not be able to see they actually have exactly zero coefficient because various contributions to them end up cancelling out; the MLDE approach on the other hand, quickly rules out those powers of $\tilde{q}.$ As another example, the MLDE approach to the $\mathrm{USp(4)}$ theory with a half-hyper in $\mathbf{16}$ naively suggests certain $\log^3 \tilde{q}$ terms in the high-temperature expansion; the direct matrix-integral approach on the other hand immediately rules out powers of $\log \tilde{q}$ higher than the rank, in this case 2. See Section~\ref{subsec:ExSec2} for more details.

Another aspect of the said interplay concerns the possibility of predicting the roots of the MLDE indicial equation from the matrix integral approach. The indicial roots dictate, modulo positive integers (or half-integers in the twisted case), the powers of $\tilde{q}$ appearing in the high-temperature expansion. These powers are determined on the other hand via the matrix integral approach as the critical values (in a sense to be clarified in Section~\ref{sec:tauTo0}) of the higher Rains functions. We identify a universally present critical value, at the origin of the BPS moduli space, corresponding to the power $2(\mathrm{a}-\mathrm{c})$ for $\tilde{q}$, with $\mathrm{a}, \mathrm{c}$ the 4d central charges. Thereby we demonstrate that there should always naturally\footnote{Here by ``naturally'' we mean absent completely destructive interferences (\emph{i.e.} cancellations) that are not anticipated based on symmetries. For an example where cancellations can be anticipated from symmetries consider a case where the Schur index has a ``$T$ symmetry'' $q\to q\, e^{2\pi i}$, so that half-integer powers of $q$ in the $q$ expansion are ruled out. This, together with modular ``$S$ symmetry'', could imply absence of certain powers of $\tilde{q}$ in the high-temperature expansion. This type of symmetry-based consideration \emph{naturally} explains certain cancellations in the high-temperature expansion of the Schur index of $\mathrm{SU}(2)$ SQCD for instance; see Sections~\ref{subsec:ExSec2} and \ref{subsec:cancelDisc}. We are not aware of analogous symmetries that could be responsible for cancelation of the $\tilde{q}^{2(\mathrm{a}-\mathrm{c})}$ terms in the high-temperature expansion, so we expect the $\tilde{q}^{2(\mathrm{a}-\mathrm{c})}$ term to be naturally present.} be a root of the indicial equation that either coincides with $2(\mathrm{a}-\mathrm{c})$, or is less than it by a positive integer (or half-integer in the twisted case). In all examples that we study, there is a root that indeed coincides with $2(\mathrm{a}-\mathrm{c})$. Often this is the smallest root, hence setting the leading asymptotic to be
\begin{equation}
    \mathcal{I}(q\to1)\approx e^{-\frac{2\pi i}{\tau}2(\mathrm{a}-\mathrm{c})}.\label{eq:DKformula}
\end{equation}
In one example, the $\mathrm{USp(4)}$ with a half-hyper in the $\mathbf{16}$, we find that there is a smaller critical value of the Rains function and hence \emph{the above leading asymptotic is modified}. From the MLDE perspective this is because there is an indicial root smaller than $2(\mathrm{a}-\mathrm{c})$ in this case. The critical value corresponding to $2(\mathrm{a}-\mathrm{c})$ is again naturally present, though in this case it corresponds to the next-to-smallest indicial root and sets the exponentially small correction to the leading asymptotic.

While the modular $S$ transformation allows probing the high-temperature expansion, more general modular transformations allow probing expansions for $q$ near roots of unity. The root-of-unity limits of 4d superconformal indices have been studied recently in \cite{ArabiArdehali:2021nsx,Jejjala:2021hlt,Cabo-Bizet:2021plf,Eleftheriou:2022kkv}. In Section~\ref{sec:tauToRat} we further develop the results of \cite{Eleftheriou:2022kkv} for the Schur index to incorporate also the exponentially suppressed terms in the expansion.
We clarify that the structure of the expansion for $q$ near an arbitrary root of unity is either similar to that around $\tau\to0$, or to the one around $\tau\to 1$ (since the Schur index might contain half-integer powers of $q$ these two limits are not equivalent in general). Just as the structure of the first expansion is governed by the Rains function and its higher analogs, the second expansion (on the ``second sheet'') is governed by what we call the \emph{twisted Rains function} and its higher analogs.

That the $\tau\to 0$ and $\tau\to 1$ expansions exhaust the structures that govern the expansions around general roots of unity can be understood easily from the MLDE perspective. In that approach there are only two inequivalent ways of taking the high-temperature limit: $I)$ by doing an $S$-transformation of the vvmf $q\to \tilde{q}:=\exp(-2\pi i/\t)$ and then taking the $\tilde{q}\to 0$ limit; $II)$ by doing first a $T$-transformation and then repeating step $I$. The latter transformation acts non-trivially only if the index involves half-integer powers of $q$ and thus square-root branch cuts, in which case $T$ takes the index to its second sheet. This perspective also makes it obvious that the first and second sheet limits coincide \emph{iff} the Schur index satisfies an untiwsted MLDE. In that case the vvmf is $T$-invariant and thus the limits $I$ and $II$ above coincide.\\

    

The rest of this paper is organized as follows. In Section~\ref{sec:tauTo0} we present a general analysis of the small-$\tau$ expansion of Schur indices, first for Lagrangian SCFTs from the matrix integral perspective and then more generally from the MLDE perspective. We then exemplify our analysis by studying four concrete cases: $\mathrm{SU}(2)$ $\mathcal{N}=4$ theory, $\mathrm{SU}(2)$ SQCD, a $\mathrm{USp}(4)$ gauge theory that violates the formula~\eqref{eq:DKformula}, and an $\mathrm{SU}(2)\times\mathrm{SU}(2)$ quiver gauge theory. In all four cases we study the leading asymptotic as well as the first exponentially suppressed correction. Section~\ref{sec:tauToRat} contains an analogous study of the small-$\tilde{\tau}$ expansion, with $\tilde{\tau}:=c\tau+d$, where $c,d$ are relatively prime integers. Section~\ref{sec:Discussion} contains a conceptual summary of our results, as well as a discussion of two interesting related problems for future investigation. Some of the necessary mathematical background for our work is summarized in the two appendices.



\section{The $\tau\to0$
expansion}\label{sec:tauTo0}

In this section we define $\tilde{q}$ via 
\begin{equation}
    \tilde{q}:=e^{-2\pi i/\tau}.
\end{equation}

\subsection{Matrix-integral perspective}

This subsection contains a general analysis of the high-temperature expansion of the Schur index using its integral expression available for Lagrangian theories. We review known results and also add new insights, in particular on the exponentially small corrections to the leading asymptotics as $\beta\propto |\tau|\to 0$.


The integral expression for the Schur index \cite{Gadde:2011ik} of a Lagrangian 4d $\mathcal{N}=2$ gauge theory with hypermultiplets of $\mathcal{N}=1$ R-charge 2/3 (the canonical value), reads (\emph{cf.}~\cite{Dolan:2008qi})  
\begin{equation}
    \mathcal{I}(q)=(q;q)^{2r_G}\int_{\mathfrak{h}_{\text{cl}}}  \frac{\mathrm{d}^{r_G}x}{|W|}\prod_{\alpha_+}\left(\frac{\Gamma_e(qz^{\pm\alpha_+})}{\Gamma_e(z^{\pm\alpha_+})}\right)\prod_{\rho^\chi} \Gamma_e(q^{1/2}z^{\rho^\chi}).\label{eq:integralSchur}
\end{equation}
Here $r_G$ is the rank of the gauge group $G$, while $|W|$ stands for the order of the Weyl group of $G$, and $\alpha_+$ are the positive roots of $G$. The weights $\rho^\chi$ go over all the weights of the gauge group representations of the chiral multiplets inside $\mathcal{N}=2$ hypermultiplets; note that we are not including among $\rho^\chi$ the weights of the chiral multiplets of $\mathcal{N}=2$ vector multiplets! We also have $z_j=e^{2\pi i x_j}$, and the symbol $z^\rho$ stands for $z_1^{\rho_1}\cdots z_{r_G}^{\rho_{r_G}}.$ The range of integration $\mathfrak{h}_{\text{cl}}$ is $-1/2<x_j<1/2$, while $\Gamma_e$ stands for the elliptic gamma function, and $(q;q)$ for the $q$-Pochhammer symbol. See~Appendix~\ref{app:special} for the definitions of the special functions $\Gamma_e$ and $(x;q)$, as well as the $\theta_0$ function used below.

Using the special function identities in \eqref{eq:simpGamThet} we can eliminate the elliptic gamma functions and  write the Schur index alternatively in terms of the theta function $\theta_0$ as
\begin{equation}
\mathcal{I}(q)=\frac{(q;q)^{2r_G}}{\theta_0(q^{1/2};q)^{n_{\rho_0}/2}}\int_{\mathfrak{h}_{\text{cl}}} \frac{\mathrm{d}^{r_G}x}{|W|}\ \frac{\prod_{\alpha_+}\theta_0(z^{\pm\alpha_+};q)}{\prod_{\rho_+^\chi}\theta_0(q^{1/2}z^{\rho^\chi_+};q)}.\label{eq:indexTheta0}
\end{equation}
Here $n^{}_{\rho_0}$ is the number of zero weights in the chiral multiplets inside $\mathcal{N}=2$ hypermultiplets.

For example, the Schur index of the USp($4$) theory with a half hyper in the $\mathbf{16}$, which will be analyzed below, can be written as
\begin{equation}
    \mathcal{I}(q)=\frac{(q;q)^{4}}{8}\oint \frac{\mathrm{d}z_1}{2\pi iz_1}\frac{\mathrm{d}z_2} {2\pi iz_2}\ \frac{\theta_0(z_1^{\pm 2},z_2^{\pm 2},(z_1z_2)^{\pm1},(z_1/z_2)^{\pm1};q)}{\theta_0(q^{1/2}z_1^2 z_2^{\pm1},q^{1/2}z_1 z_2^{\pm2},q^{1/2}z_1,q^{1/2}z_1,q^{1/2}z_2,q^{1/2}z_2;q)},
\end{equation}
with the unit-circle contour for $z_{1,2}$. This is obtained by substituting into \eqref{eq:indexTheta0} the set of positive roots of USp($4$) and the positive weights of its $\mathbf{16}$ representation,
\begin{equation}
	\alpha_+ = \{ (2,0),(0,2),(1,1),(1,-1)¬†\} ,¬†\;\;\rho^\chi_+ = \{ (2,1),(2,-1),(1,2),(1,-2),(1,0),(0,1)¬†\},
\end{equation}
where the last two elements of $\rho^\chi_+$ have multiplicity two.

To obtain the high-temperature expansion of \eqref{eq:indexTheta0}, we use the $S$-transformed expression for the $q$-Pochhammer symbol
\begin{equation}
    (q;q)^2 \; = \;  
    \frac{1}{{-i\,\tau}} \;  \exp \Bigl( -\frac{2\pi i}{12\, \tau}  -\frac{2\pi i\tau}{12}\Bigr)\, \prod_{k=1}^\infty (1-\tilde{q}^k)^2,
\label{eq:PochEst}
\end{equation}
where $\tilde{q}=e^{-2\pi i/\tau},$ as well as
\begin{equation}
    \theta_0(\tau/2;\tau)=2\,e^{\frac{i\pi\tau}{12}-\frac{i\pi}{6\tau}}\prod_{k=1}^\infty (1+\tilde{q}^k)^2,\label{eq:theta0t/2Smain}
\end{equation}
and \eqref{eq:theta0S}  to get
\begin{equation}
  \begin{split}
  \mathcal{I}(q)&=e^{-i\pi \tau \mathrm{c}}\, \frac{1}{(-i\tau)^{r_G}}\, e^{-\frac{2\pi i}{\tau}2(\mathrm{a}-\mathrm{c})}\times \frac{1}{2^{n_{\rho_0}/2}}\, \bigg(\prod_{k=1}^\infty \frac{(1-\tilde{q}^k)^{2r_G}}{(1+\tilde{q}^k)^{n^{}_{\rho_0}}} \bigg)\times\\
  &\ \  \int_{\mathfrak{h}_{\text{cl}}} \frac{\mathrm{d}^{r_G}x}{|W|}\, e^{-\frac{2\pi i}{\tau}\cdot L(\boldsymbol{x})} \ \prod_{l=1}^\infty\, \frac{\prod_{\alpha_+} (1-\tilde{q}^l e^{\frac{2\pi i}{\tau}(\{\alpha_+\cdot\boldsymbol{x}\})})^2\, (1-\tilde{q}^l  e^{\frac{2\pi i}{\tau}(1-\{\alpha_+\cdot\boldsymbol{x}\})})^2}{\prod_{\rho^\chi_+} (1+\tilde{q}^l  e^{\frac{2\pi i}{\tau}(\{\rho^{\chi}_+\cdot\boldsymbol{x}\})})(1+\tilde{q}^l  e^{\frac{2\pi i}{\tau}(1-\{\rho^{\chi}_+\cdot\boldsymbol{x}\})})}.
    \end{split}\label{eq:indexStransformed}
\end{equation}
Here we have used
\begin{equation}
    \begin{split}
        \mathrm{c}&=\mathrm{dim}G/6+\sum_\chi\mathrm{dim}\mathcal{R}_\chi/24,\\
        \mathrm{a}&=5\,\mathrm{dim}G/24+\sum_\chi\mathrm{dim}\mathcal{R}_\chi/48.
    \end{split}\label{eq:LagrangianCentralCharges}
\end{equation}
Note that the $\mathcal{N}=1$ $U(1)_R$ 't~Hooft anomaly is $\mathrm{Tr}R=2\,\mathrm{dim}G/3-\sum_\chi\mathrm{dim}\mathcal{R}_\chi/3\,.
    $

In \eqref{eq:indexStransformed} the function $L(\boldsymbol{x})$ is the $\mathcal{N}=2$ Rains function in the Schur limit \cite{ArabiArdehali:2015ybk}
\begin{equation}
    L(\boldsymbol{x})=\frac{3}{2}L_h(\boldsymbol{x})=\frac{1}{2}\sum_{\rho_+^\chi}\vartheta(\rho_+^\chi\cdot\boldsymbol{x})-\sum_{\alpha_+}\vartheta(\alpha_+\cdot\boldsymbol{x}), 
\end{equation}
\begin{equation}
\vartheta(x):=\{x\}\big(1-\{x\}\big),
\end{equation}
with $\{x\}:=x-\lfloor x\rfloor.$ In this work we will refer to $L(\boldsymbol{x})$ as the Rains function.\footnote{Beware that in \cite{ArabiArdehali:2015ybk} it was $L_h(\boldsymbol{x})=2L(\boldsymbol{x})/3$ that was called the Rains function.} It is a piecewise linear function thanks to the $\mathcal{N}=1$ $U(1)_R$ ABJ anomaly cancelation \cite{ArabiArdehali:2015ybk}.

Now we proceed to the high-temperature expansion. The main difficulty is that the naive Taylor expansion 
\begin{equation}
    \frac{1}{1+X}=1-X+X^2\mp\cdots,\label{eq:naiveTaylor}
\end{equation}
can not be applied to the $l=1$ factors
\begin{equation}
    \frac{1}{\prod_{\rho^\chi_+} (1+\tilde{q}  e^{\frac{2\pi i}{\tau}(\{\rho^{\chi}_+\cdot\boldsymbol{x}\})})(1+\tilde{q}  e^{\frac{2\pi i}{\tau}(1-\{\rho^{\chi}_+\cdot\boldsymbol{x}\})})},
\end{equation}
inside the integrand of \eqref{eq:indexStransformed}. The reason is that as $\{\rho^\chi_+\cdot\boldsymbol{x}\}$ approaches $0$ (respectively $1$), we have $\tilde{q}  e^{\frac{2\pi i}{\tau}(1-\{\rho^{\chi}_+\cdot\boldsymbol{x}\})}$ (respectively $\tilde{q}  e^{\frac{2\pi i}{\tau}(\{\rho^{\chi}_+\cdot\boldsymbol{x}\})}$) approaching the boundary $|X|=1$ of the domain of convergence of the series \eqref{eq:naiveTaylor}. As in \cite{Ardehali:2021irq}, we confront this difficulty by decomposing the integration domain.\\

\noindent\textbf{Terminology.} Here we decompose the domain of integration into an ``outer patch'' $\mathcal{S}'_\epsilon$, where all $\rho^\chi_+\cdot \boldsymbol{x}$ are at least $\epsilon$ away from integer values, and finitely many ``inner patches'' that complement the outer patch. (Note that the outer patch is the disjoint union of finitely many convex polytopes $\mathcal{P}_m^\epsilon$---see the six blue triangles in Figure~\ref{fig:su3decomp}.) Then, in the integrand of \eqref{eq:indexStransformed}, use uniformly convergent small-$\tilde{q}$ expansions appropriate to each patch. We will refer to the hyperplanes where $\rho^\chi_+\cdot \boldsymbol{x}$ is exactly $\epsilon$ away from an integer as ``cut-off surfaces'', to the subspaces where $\rho^\chi_+\cdot \boldsymbol{x}\in\mathbb{Z}$ as ``singular hyperplanes'', and to the set $\mathcal{S}$ of all singular hyperplanes as the ``singular set''.

\subsubsection{The outer patch}\label{subsubsec:outer}

In the outer patch we can Taylor expand all the factors in the denominator of the integrand of \eqref{eq:indexStransformed}, and rewrite the contribution to the index as
\begin{equation}
   \begin{split}
   \mathcal{I}_{\text{out}}(q)
&= e^{-i\pi \tau \mathrm{c}} \, e^{-\frac{2\pi i}{\tau}2(\mathrm{a}-\mathrm{c})}\bigg(\int_{\mathcal{S}'_\epsilon} \frac{\mathrm{d}^{r_G}\boldsymbol{x}}{(-i\tau)^{r_G}}\ C_1\  e^{-\frac{2\pi i}{\tau}\cdot L(\boldsymbol{x})}+\int_{\mathcal{S}'_\epsilon} \frac{\mathrm{d}^{r_G}\boldsymbol{x}}{(-i\tau)^{r_G}}\  C_2\ e^{-\frac{2\pi i}{\tau}\cdot L^{(2)}(\boldsymbol{x})}+\cdots\\
&\qquad\qquad\qquad \qquad \qquad +\int_{\mathcal{S}'_\epsilon} \frac{\mathrm{d}^{r_G}\boldsymbol{x}}{(-i\tau)^{r_G}}\  C_n\ e^{-\frac{2\pi i}{\tau}\cdot L^{(n)}(\boldsymbol{x})}+\cdots\bigg),
\end{split}\label{eq:outExpansion}
\end{equation}
with $C_n$ constants (\emph{i.e.} independent of $\tau,x_j$). We refer to the function $L^{(n)}$ above as the $n$th Rains function. As it is clear from \eqref{eq:indexStransformed}, we have
\begin{equation} 
L^{(n)}(\boldsymbol{x})=L(\boldsymbol{x})+\mathbb{Z}_{\ge0}\big(\{\alpha_+\cdot \boldsymbol{x}\},1-\{\alpha_+\cdot \boldsymbol{x}\},\{\rho^\chi_+\cdot \boldsymbol{x}\},1-\{\rho^\chi_+\cdot \boldsymbol{x}\},1\big).\label{eq:higherRains}
\end{equation}
That is, the $n$th Rains function is the (1st) Rains function plus a linear combination with non-negative integer coefficients of the ``generators'' shown above. Note in particular that $L^{(n)}$ is piecewise linear.

We have not ordered $L^{(n)}$ according to any specific scheme. It should be clear though that no higher Rains function has a minimum less than that of the (1st) Rains function $L$, hence the leading asymptotic is given by the first term on the RHS of \eqref{eq:outExpansion}. 

However, as far as the general structure of the integral goes, all the terms are similar. Since all $L^{(n)}$, including $L^{(1)}:=L$ are piecewise linear, their integral over the outer patch decomposes into integrals over finitely many polytopes $\mathcal{P}_m^\epsilon$ over which the Rains functions are linear \cite{Ardehali:2021irq}. As shown in \cite{Barvinok1992ComputingTV}, integral of the exponential of a linear function over a convex polytope is simply given by a sum over the ``critical sets'' of the linear function, defined as the $j$-faces of the polytope on which the linear function is constant. (Note that all vertices of the polytope are critical sets.) Each critical set, denoted $\ast$, contributes to the $n$th integral an expression of the form
\begin{equation}
    \Delta\mathcal{I}_\text{out}^\ast(q)= e^{-\frac{2\pi i}{ \tau}(2(\mathrm{a}-\mathrm{c})+L_\ast^{(n)})}\cdot \frac{C_\ast}{(-i\tau)^{\mathrm{dim}\mathfrak{h}^\ast}}\cdot e^{-i\pi \tau \mathrm{c}},\label{eq:DeltaIast}
\end{equation}
where $C_\ast$ is some constant (\emph{i.e.} independent of $\tau$), while $\mathrm{dim}\mathfrak{h}^\ast$ and $L^{(n)}_\ast$ are the dimension of the critical set and the value of $L^{(n)}$ on it. We refer to $L^{(n)}_\ast$ as the ``critical values'' of the higher Rains functions.

Summing over all the critical sets in the outer patch \eqref{eq:DeltaIast} yields
\begin{equation}
    {q^{\mathrm{c}/2}\,\mathcal{I}_\text{out}(q)=\tilde{q}^{h_{\text{min}}} \cdot C_0(\log\tilde{q})^{\mathrm{dim}\mathfrak{h}_{\mathrm{qu}}}+\sum_\ast \tilde{q}^{h_\ast} \cdot C_\ast(\log\tilde{q})^{\mathrm{dim}\mathfrak{h}^\ast},}\label{eq:tauTo0Iout}
\end{equation}
where
\begin{equation}
    \begin{split}
        h_{\text{min}}=2(\mathrm{a}-\mathrm{c})+L^{\mathrm{min}},\qquad h_\ast=2(\mathrm{a}-\mathrm{c})+L^{(n)}_{\ast}.
    \end{split}
\end{equation}

\subsubsection{Inner patches}\label{subsubsec:inner}

The analysis of the inner patches, although technically more involved, is not expected to modify much the structure of the answer \eqref{eq:tauTo0Iout}. The main difference will be that the monomials $C_0(\log\tilde{q})^{\mathrm{dim}\mathfrak{h}_{\mathrm{qu}}}$ and $C_\ast(\log\tilde{q})^{\mathrm{dim}\mathfrak{h}^\ast}$ will become more general polynomials of the same degree in $\log \tilde{q}$, as in Eq.~\eqref{eq:tauTo0FullAllExpSmall}). This claim will be justified below. The reader not interested in the technical details is advised to skip to Eq.~\eqref{eq:tauTo0FullAllExpSmall}.\footnote{As \cite{Polchinski:2017vik} emphasized in a related context  ``once you say to work scale-by-scale,
the rest is just bookkeeping.'' }

Denote the $m$th inner patch by in$_m$. Define the light weights $\rho_+^L$ associated to it as those for which $\rho^\chi_+\cdot \boldsymbol{x}$ is \emph{not} at least $\epsilon$ away from integers inside in$_m$.
The corresponding $l=1$ factors in the integrand of the index~\eqref{eq:indexStransformed} are not uniformly small everywhere. Rather than incorporating them in the expansion, then, we keep those factors in the integrand. This leads to a non-exponential factor of the form
\begin{equation}
    f_m(\boldsymbol{x},\tau):=\frac{1}{\prod_{\rho_+^L }(1+\tilde{q}e^{\frac{2\pi i}{\tau}(\{\pm\rho_+^L\cdot\boldsymbol{x}\})})}.
\end{equation}

The $l>1$ factors of the light weights, as well as all $l\ge1$ factors of the rest of weights and roots will be expanded in exponentials as in \eqref{eq:outExpansion}. This yields exponentials of piecewise linear functions $L_{\text{in}_m}^{(n)}(\boldsymbol{x})$ that we call the $n$th modified Rains functions. We can hence write
\begin{equation}
   \begin{split}
   \mathcal{I}_{\text{in}_m}(q)
&= e^{-i\pi \tau \mathrm{c}} \, e^{-\frac{2\pi i}{\tau}2(\mathrm{a}-\mathrm{c})}\bigg(\int_{\text{in}_m} \frac{\mathrm{d}^{r_G}\boldsymbol{x}}{(-i\tau)^{r_G}}\ \tilde{C}_1\ f_m(\boldsymbol{x},\tau) \   e^{-\frac{2\pi i}{\tau}\cdot L_{\text{in}^{}_m}(\boldsymbol{x})}+\cdots\\
&\qquad\qquad\qquad \qquad \qquad \qquad +\int_{\text{in}_m} \frac{\mathrm{d}^{r_G}\boldsymbol{x}}{(-i\tau)^{r_G}}\  \tilde{C}_n\ f_m(\boldsymbol{x},\tau)\  e^{-\frac{2\pi i}{\tau}\cdot L_{\text{in}_m}^{(n)}(\boldsymbol{x})}+\cdots\bigg),
\end{split}\label{eq:inExpansion}
\end{equation}
with $\tilde{C}_n$ constants (\emph{i.e.} independent of $\tau,x_j$). The generators of $L^{(n)}$ which are  missing in $L_{\text{in}_m}^{(n)}$, namely $\{\pm\rho_+^L\cdot\boldsymbol{x}\}$, are contained in the function $f_m(\boldsymbol{x},\tau)$. As we will see they reappear in the final answer---see Eq.~\eqref{eq:tauTo0FullAllExpSmall}.\\

\subsubsection*{Without singular intersections: integrals over generic prismatoids}\label{subsec:Fuchsian}

We first assume in$_m$ contains (parts of) a number of hyperplanes $\mathcal S_1,\dots,\mathcal S_p,\dots$ belonging to the singular set $\mathcal S$, but \emph{does not contain any intersections} of those hyperplanes. That is, $\mathrm{in}_m\cap\mathcal{S}$ consists of $\mathrm{in}_m\cap\mathcal{S}_1,\dots,\mathrm{in}_m\cap\mathcal{S}_p,\dots$ that do not intersect, and are hence at nonzero distances of order $\epsilon$ from each other.

We organize the analysis in four steps.

\paragraph{Step 1: decomposing the integration domain to simple building blocks.} Choose $\epsilon_1$ small enough such that an $\epsilon_1$ neighborhood of $\mathcal{S}_p$ inside in$_m$, denoted $\text{in}_m\cap\mathcal{S}^{\epsilon_1}_p$, does not intersect any other component of $\text{in}_m\cap\mathcal{S}$. Then, inside in$_m$ and outside such $\epsilon_1$ neighborhoods $f_m$ can be expanded in exponentials and we are back to integrals of the form in \eqref{eq:outExpansion}, so the contribution of such regions to the index is again of the form \eqref{eq:tauTo0Iout}.

Now consider the contribution from $\text{in}_m\cap\mathcal{S}^{\epsilon_1}_p$. Note that $\text{in}_m\cap\mathcal{S}^{\epsilon_1}_p$ is a prismatoid with the parallel faces both of co-dimension one. We refer to such prismatoids as \emph{generic}.

On $\text{in}_m\cap\mathcal{S}^{\epsilon_1}_p$, we expand the denominators in exponentials for all the terms except the ones corresponding to the light weights of $\mathcal{S}_p$. Then we would be dealing with integrals of the form
\begin{equation}
    \int_{\mathrm{in}_m\cap\mathcal{S}^{\epsilon_1}_p} \frac{\mathrm{d}^{r_G}\boldsymbol{x}}{(-i\tau)^{r_G}} \,\tilde{C} \,
    \frac{1}{\prod_{\rho_+^L }(1+e^{-\frac{2\pi i}{\tau}(\{\pm\rho_+^L\cdot\boldsymbol{x}\})})}\ e^{-\frac{2\pi i}{\tau}L_p^{(n)}(\boldsymbol{x})}.\label{eq:InStep1Int}
\end{equation}

\paragraph{Step 2: linear decompositions in the integrand and rescaling.} We parametrize $\mathrm{in}_m\cap\mathcal{S}^{\epsilon_1}_p$, which is a prismatoid, by $\boldsymbol{x}_{||}$ parallel to $\mathcal{S}_p$, and $x_\perp$ perpendicular to it. Since $L^{(n)}_p$ is linear, it can be written as a sum of linear functions of $\boldsymbol{x}_{||}$ and $x_\perp$:
\begin{equation}
L^{(n)}_p(\boldsymbol{x})=f(\boldsymbol{x}_{||})+a_0\label{eq:f||Def} x_{\perp},
\end{equation}
with some constant $a_0.$ Linearity further implies $f(\boldsymbol{x}_{||})=f(\boldsymbol{0})+\tilde{f}(\boldsymbol{x}_{||})$, with $\tilde{f}$ homogeneous.

As a corollary, on a vertex $v$ of a constant-$x_\perp$ section of the prismatoid, linearity of $\boldsymbol{x}_{||}^v$ in $x_\perp$, together with \eqref{eq:f||Def} which allows fixing the value at $x_\perp=0,$ imply that $f\big(\boldsymbol{x}_{||}^v\big)=a_1x_\perp+L_p^{(n)}\big(\boldsymbol{x}_{||}^v(x_\perp=0),x_\perp=0\big)$, with some constant $a_1$. Thus we can write
\begin{equation}
L^{(n)}_p(\boldsymbol{x}^v)=L_p^{(n)}\big(\boldsymbol{x}_{||}^v(x_\perp=0),x_\perp=0\big)+ax_{\perp},\label{eq:LpVertex}
\end{equation}
where $a:=a_0+a_1$. 
With a rescaling of the integration variable $2\pi \boldsymbol{x}/(-i\tau)\to \boldsymbol{x}$, the integral \eqref{eq:InStep1Int} takes the form
\begin{equation}
    e^{-\frac{2\pi i}{\tau}f(\boldsymbol{0})}\int_{\frac{2\pi}{-i\tau}(\mathrm{in}_m\cap\mathcal{S}^{\epsilon_1}_p)} \mathrm{d}^{r_G}\boldsymbol{x} \ \tilde{C} \,
    \frac{1}{\prod_{\rho_+^L }(1+e^{-(\{\pm\rho_+^L\cdot\boldsymbol{x}\})})}\ e^{-(\tilde{f}(\boldsymbol{x}_{||})+a_0 x_\perp)}.\label{eq:InStep2Int}
\end{equation}

\paragraph{Step 3: reduction to a single variable.} At fixed $x_\perp$, the integral over $\boldsymbol{x}_{||}$ is of the exponential type that we already know how to deal with. First assume the resulting integral will localize to the vertices of the constant-$x_\perp$ sections of the prismatoid, yielding exponentials of the form $e^{-\tilde{f}(\boldsymbol{x}_{||}^v)}$, with $v$ denoting the vertices. Using the expression for $f(\boldsymbol{x}_{||}^v)$ above \eqref{eq:LpVertex}, the remaining integrals over $x_\perp$ will be of the form
\begin{equation}
    e^{-\frac{2\pi i}{\tau}L_p^{(n)}\big(\boldsymbol{x}_{||}^v(x_\perp=0),x_\perp=0\big)}\int_0^{\frac{2\pi\epsilon_1}{-i\tau}} \mathrm{d}x_\perp \ \tilde{C} \,
    \frac{1}{\prod_{\rho_+^L }(1+e^{-(\{\pm\rho_+^L\cdot\boldsymbol{x}\})})}\ e^{-a x_\perp}.\label{eq:simplifiedInner0}
\end{equation}

\paragraph{Step 4: Renormalization Group Equation.} The integral above can be written as
\begin{equation}
    I(\Lambda)=\int_0^\Lambda \mathrm{d}x_\perp \tilde{C}\frac{1}{\prod_i (1+e^{-n_i  x_\perp})^{m_i}}e^{-a x_\perp},\label{eq:LambdaInt1}
\end{equation}
where $n_i,m_i$ are integers, and $\Lambda=2\pi\epsilon_1/(-i\tau)$.

We can not Taylor-expand the fraction in the integrand of \eqref{eq:LambdaInt1} because of the breakdown of uniform convergence near $x_\perp=0$. We overcome this difficulty by considering ${\mathrm{d}}I(\Lambda)/{\mathrm{d}\Lambda}$:
\begin{equation}
    \frac{\mathrm{d}}{\mathrm{d}\Lambda}I(\Lambda)=\tilde{C} \frac{1}{\prod_i (1+e^{-n_i \Lambda})^{m_i}}e^{-a \Lambda}.\label{eq:rgODE}
\end{equation}
The RHS of this RG-type equation (RGE) can now be expanded in powers of $e^{-\Lambda}$. Then, we integrate the two sides with respect to $\Lambda$. Except for determination of the integration constant,  which should be done using variations of the subtraction method (see footnote~\ref{ftnt:subtractionMethod}), the rest of the calculation is completely straightforward. 

The integration constant carries in particular the contribution from $\mathcal{S}_p$ (one end of the prismatoid), while the result of the indefinite integral carries the contribution from the $x_\perp=\epsilon_1$ hyperplane parallel to $\mathcal{S}_p$ (the other end of the prismatoid). If $a\neq0$, the latter is of the form
\begin{equation}
    e^{-a\Lambda}\sum_i\big(c'_{0i}+c'_{1i} e^{-n_i \Lambda}+c'_{2i} e^{-2n_i \Lambda}+\dots\big).\label{eq:2ndFuchsSol}
\end{equation}
 Notice how the missing generators discussed below~\eqref{eq:inExpansion} reappear through the higher powers of $e^{-\Lambda}.$ If $a=0$, we have instead (again modulo the integration constant)
\begin{equation}
    c'_{0}\Lambda+\sum_i\big(c'_{1i} e^{-n_i \Lambda}+c'_{2i} e^{-2n_i \Lambda}+\dots\big).\label{eq:2ndFuchsSolLog}
\end{equation}

Now let's relax our earlier assumption that the integral over $\boldsymbol{x}_{||}$ localizes to vertices; assume that it rather localizes on higher-dimensional faces of the constant-$x_\perp$ sections of the prismatoid. Then instead of \eqref{eq:simplifiedInner0} we would end up with integrals of the form
\begin{equation}
    e^{-\frac{2\pi i}{\tau}L_p^{(n)}\big(\boldsymbol{x}_{||}^f(x_\perp=0),x_\perp=0\big)}\int_0^{\epsilon_1} \frac{\mathrm{d}x_\perp}{-i\tau} \,\tilde{C} \,
   \frac{V_f(x_\perp)}{(-i\tau)^{d_f}}\, \frac{1}{\prod_{\rho_+^L }(1+e^{-\frac{2\pi i}{\tau}(\{\pm\rho_+^L\cdot\boldsymbol{x}\})})}\ e^{-\frac{2\pi i}{\tau}a x_\perp},\label{eq:simplifiedInner1}
\end{equation}
where $V_f$ is the volume of the localization faces, and $d_f$ their dimension. Since $V_f(x_\perp)$ is a polynomial of maximum degree $d_f$ in $x_\perp$, the integral in \eqref{eq:simplifiedInner1} will be a sum of integrals of the form
\begin{equation}
    \frac{1}{(-i\tau)^{d_f-d}}\int_0^{\epsilon_1} \frac{\mathrm{d}x_\perp}{-i\tau} \,\tilde{C} \,\left(
   \frac{x_\perp}{-i\tau}\right)^d\, \frac{1}{\prod_{\rho_+^L }(1+e^{-\frac{2\pi i}{\tau}(\{\pm\rho_+^L\cdot\boldsymbol{x}\})})}\ e^{-\frac{2\pi i}{\tau}a x_\perp},\label{eq:simplifiedInner2}
\end{equation}
with $d\in\{0,1,\dots,d_f\}$. We will then have integrals analogous to \eqref{eq:LambdaInt1} of the form 
\begin{equation}
    I(\Lambda)=\int_0^\Lambda \mathrm{d}x_\perp \tilde{C}\frac{x_\perp^d}{\prod_i (1+e^{-n_i  x_\perp})^{m_i}}e^{-a x_\perp},\label{eq:LambdaInt2}
\end{equation}
and the analog of the RG equation \eqref{eq:rgODE} would look like
\begin{equation}
    \frac{\mathrm{d}}{\mathrm{d}\Lambda}I(\Lambda) =\tilde{C}\frac{\Lambda^d e^{-a\Lambda}}{\prod_i (1+e^{-n_i\Lambda})^{m_i}}.\label{eq:FuchsPrepare2}
\end{equation}
Again the RHS of this can be safely expanded in powers of $e^{-\Lambda}$. Then we integrate both sides with respect to $\Lambda$. The integration constant carries in particular the contribution from $\mathcal{S}_p$ (one end of the prismatoid), while the result of the indefinite integral carries the contribution from the other end of the prismatoid. If $a\neq0$ the latter is of the form
\begin{equation}
    e^{-a\Lambda}\sum_i\big(P_{d\,0i}(\Lambda)+P_{d\, 1i}(\Lambda) e^{-n_i \Lambda}+P_{d\, 2i}(\Lambda) e^{-2n_i \Lambda}+\dots\big),\label{eq:2ndFuchsSolD}
\end{equation}
with $P_d$ standing for degree-$d$ polynomials.  If $a=0$, we get instead
\begin{equation}
    P_{d+1\,0i}(\Lambda)+\sum_i\big(P_{d\, 1i}(\Lambda) e^{-n_i \Lambda}+P_{d\, 2i}(\Lambda) e^{-2n_i \Lambda}+\dots\big).\label{eq:2ndFuchsSolDLog}
\end{equation}
Notice again how the higher powers of $e^{-\Lambda}$ in \eqref{eq:2ndFuchsSolD} and \eqref{eq:2ndFuchsSolDLog} restore the missing generators discussed below \eqref{eq:inExpansion}. Including the integration constant and also the exponential pre-factors as in \eqref{eq:simplifiedInner1} we get inner patch contributions of the form
\begin{equation}
    e^{-\frac{2\pi i}{\tau}L_p^{(n)}\big(\boldsymbol{x}_{||}^f(x_\perp=0),x_\perp=0\big)}\left(c'+e^{-a\Lambda}\sum_i\big(P_{d\,0i}(\Lambda)+P_{d\, 1i}(\Lambda) e^{-n_i \Lambda}+P_{d\, 2i}(\Lambda) e^{-2n_i \Lambda}+\dots\big)\right),\label{eq:inFinal1}
\end{equation}
or
\begin{equation}
    e^{-\frac{2\pi i}{\tau}L_p^{(n)}\big(\boldsymbol{x}_{||}^f(x_\perp=0),x_\perp=0\big)}\left(c'+P_{d+1\,0i}(\Lambda)+\sum_i\big(P_{d\, 1i}(\Lambda) e^{-n_i \Lambda}+P_{d\, 2i}(\Lambda) e^{-2n_i \Lambda}+\dots\big)\right).\label{eq:inFinal2}
\end{equation}
The $\Lambda$-dependent pieces must cancel against similar contributions from other patches (since the final answer should be independent of our decomposition scheme, and in particular the value of $\Lambda$), and the $\Lambda$-independent pieces survive in the final answer compatible with our claim in \eqref{eq:tauTo0FullAllExpSmall}.

\subsubsection*{With singular intersections: integrals over degenerate prismatoids}\label{subsec:open}

\begin{figure}[t]
\centering
    \includegraphics[scale=0.07]{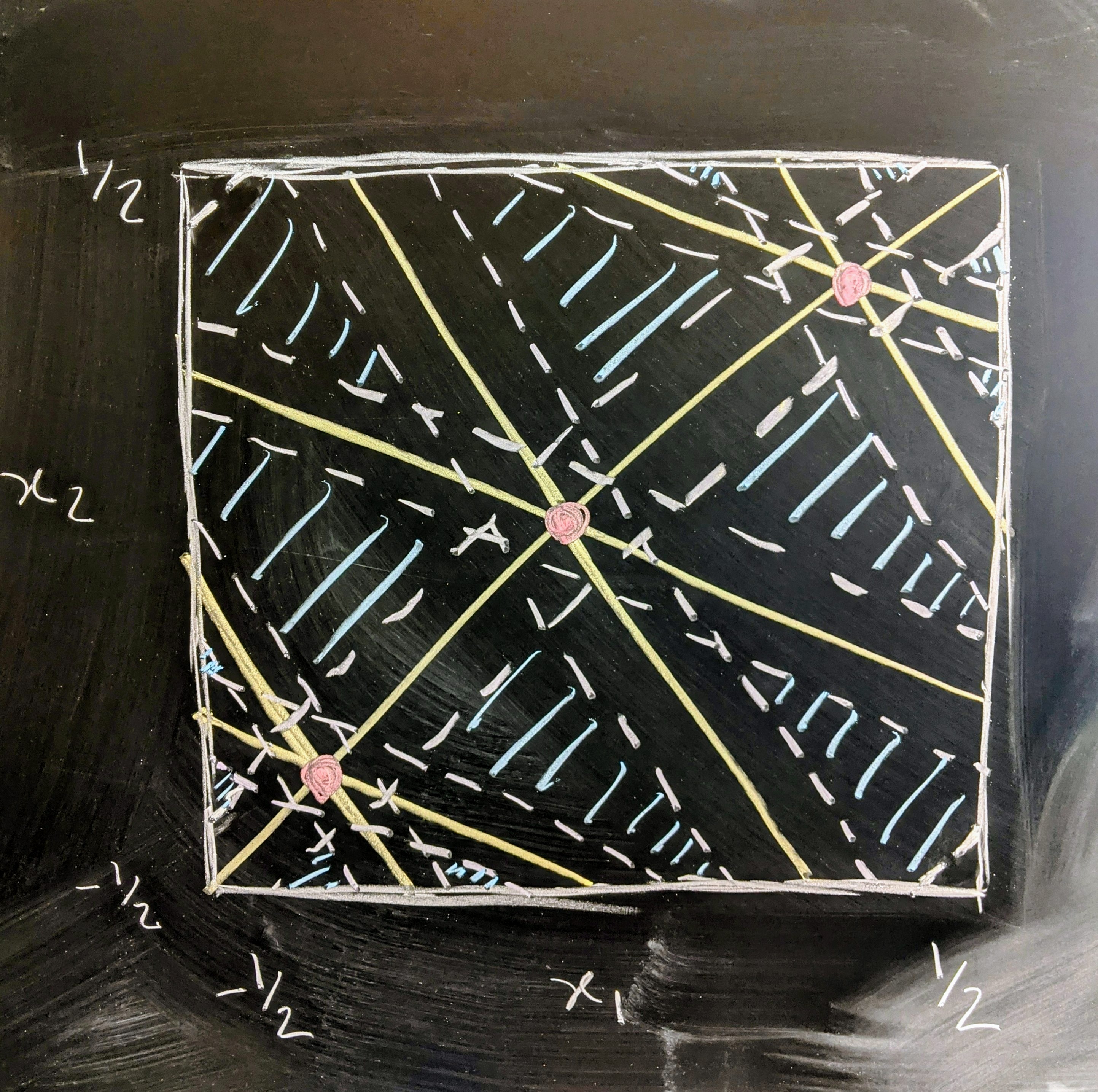}
\caption{The BPS moduli space of the $\mathrm{SU}(3)$ $\mathcal{N}=4$ theory. The red points mark singular intersections. The outer patch is crosshatched in blue. The singular set consists of the yellow lines. The dashed lines indicate the cut-off surfaces.
\label{fig:su3decomp}}
\end{figure}

See Figure~\ref{fig:su3decomp} for examples of inner patches with singular intersections.

In this case we adopt a set of new
coordinates---affinely related to $x_i$ and with unit
Jacobian---that are convenient on $\mathrm{in}_n \cap \mathcal{P}_m$. We pick a point
on the singular intersection as the new origin, and parameterize
the intersection with $\bar{x}_1,...,\bar{x}_{j_n}$. We take
$x_{\mathrm{in}}$ to parameterize a direction perpendicular to all
the $\bar{x}$s, and to increase as we go away from
the intersection and into the interior of $\mathrm{in}_n \cap\mathcal{P}_m$.
Finally, we pick $\tilde{x}_1,...,\tilde{x}_{r_G-j_n-1}$ to
parameterize the perpendicular directions to $x_{\mathrm{in}}$ and
the $\bar{x}$s. We collectively denote $\bar{x}$s and $\tilde{x}$s by $\boldsymbol{x}_{||}.$ 

Now, in an analog of Step~1 above, we decompose the integration domain and focus on the most challenging contribution which is from an $x^{}_\text{in}<\epsilon_1$ neighborhood of the intersection.

In an analog of Step~2 above, we rescale the integration variables by $2\pi/(-i\tau)$, blowing up the integration domain to a large prismatoid with the parallel faces one of co-dimension one and the other of co-dimension greater than one. We refer to such prismatoids as \emph{degenerate}. (Note that the most degenerate prismatoids in this sense are pyramids.) The integrals we are facing now are of the form
\begin{equation}
    e^{-\frac{2\pi i}{\tau}f(\boldsymbol{0})}\int_{\frac{2\pi}{-i\tau}(\mathrm{in}_m\cap\,(x_\text{in}<\epsilon_1))} \mathrm{d}^{r_G}\boldsymbol{x} \ \tilde{C} \,
    \frac{1}{\prod_{\rho_+^L }(1+e^{-(\{\pm\rho_+^L\cdot\boldsymbol{x}\})})}\ e^{-(\tilde{f}(\boldsymbol{x}_{||})+a_0 x^{}_\text{in})}.\label{eq:InStep2IntSing}
\end{equation}

Here the analysis diverges from the one on generic prismatoids: Step~3 above can not be adapted to the present case; instead we make repeated use of a variant of Step~4.

The variant consists of differentiating with respect to $\frac{1}{-i\tau}$. Since the integrand is independent of $\tau$ thanks to the re-scaling, this differentiation only ``moves'' the boundary of the integration domain. (In fact, not all boundaries are moved, but only those that correspond to cut-off surfaces.) The derivative is hence the same integral localized to such boundaries. These boundaries are \emph{less degenerate} prismatoids than the domain $\mathrm{in}_m\cap\,(x_\text{in}<\epsilon_1)$ that we started with. Hence we have made progress. Repeated differentiation of this kind localizes the integral further and further to the boundaries, which are less and less degenerate. This procedure ends when the last differentiation localizes the integral to the vertices of the boundary of $\mathrm{in}_m\cap\,(x_\text{in}<\epsilon_1)$ (or more precisely, those parts of the boundary that correspond to cut-off surfaces). Integrating the resulting RG-type equations then produces the same structure for the high-temperature expansion that we got in the case without singular intersections.

In summary, we find
\begin{equation}\label{eq:tauTo0FullAllExpSmall}
    \boxed{\begin{split}
q^{\mathrm{c}/2}\,\mathcal{I}(q)&=\tilde{q}^{h_{\text{min}}}\cdot P_{\mathrm{dim}\mathfrak{h}_\text{qu}}(\log\tilde{q}) +\sum_{n=1}^\infty\sum_{\ast_n} \tilde{q}^{h_{\ast_n}} \cdot P_{\mathrm{dim}\mathfrak{h}^{\ast_n}}(\log\tilde{q})\\
&=e^{-\frac{2\pi i}{\tau}[2(\mathrm{a}-\mathrm{c})+L^{\mathrm{min}}]}\cdot P_{\mathrm{dim}\mathfrak{h}_\text{qu}}(\log\tilde{q}) +\sum_{n=1}^\infty\sum_{\ast_n} e^{-\frac{2\pi i}{\tau}[2(\mathrm{a}-\mathrm{c})+L_\ast^{(n)}]} \cdot P_{\mathrm{dim}\mathfrak{h}^{\ast_n}}(\log\tilde{q}),
\end{split}}
\end{equation}
where $P_{\mathrm{dim}\mathfrak{h}^{\ast_n}}(\log\tilde{q})$ is a polynomial of order ${\mathrm{dim}\mathfrak{h}^{\ast_n}}$ in $\log\tilde{q}$.

Since the end result \eqref{eq:tauTo0FullAllExpSmall} is independent of $\epsilon,$ we can send $\epsilon\to0$ and observe that $L^{(n)}_\ast$ can be considered the value of the $n$th Rains function on the vertices of the polytopes $\mathcal{P}_m:=\lim_{\epsilon\to0}\mathcal{P}^\epsilon_m$. In other words, $h_{\ast_n}$ can be found by evaluating $L^{(n)}$ on the vertices of the polytopes $\mathcal{P}_m$. We will refer to these as ``vertex values'' of $L^{(n)}$.

\noindent\textbf{Remark 1.} We observe that the origin is always a critical set of the Rains function, with value $L_\ast=0$. This follows from noting that near $x=0$ we can replace $\vartheta(x)$ with\footnote{First we use $\vartheta(x)=|x|-x^2$ valid for $x\in[-1,1]$, and then since the quadratic terms cancel in the Rains function we drop the $x^2$.} $|x|$ to deduce that the Rains function simplifies to \cite{ArabiArdehali:2015ybk}
\begin{equation}
    \tilde{L}(\boldsymbol{x})=\frac{1}{2}\sum_{\rho_+^\chi}|\rho_+^\chi\cdot\boldsymbol{x}|-\sum_{\alpha_+}|\alpha_+\cdot\boldsymbol{x}|.\label{eq:Ltilde}
\end{equation}
Therefore the expansion in \eqref{eq:tauTo0FullAllExpSmall} always  contains a term corresponding to 
\begin{equation}\label{hstarOrig}
    h_\ast^{\mathrm{origin}}=2(\mathrm{a}-\mathrm{c}).
\end{equation}

\noindent\textbf{Remark 2.} There are finitely many vertices of $\mathcal{P}_m$, hence finitely many vertex values of the (1st) Rains function. The vertex values of the higher Rains functions are the same up to linear combinations with non-negative integer coefficients of the vertex values of the generators. The latter are easily seen to be rational numbers (because the Cartesian coordinates of the vertices on $\mathfrak{h}_\text{cl}$ are rational). Therefore, after getting multiplied by integers, they only provide finitely many more rational numbers up to integers. This establishes that the set of vertex values of the Rains functions, and hence the set of $h^{}_{\ast_n}$, is finite up to addition of non-negative integers.

\subsection{MLDE perspective}\label{subsec:lmde}

To any four-dimensional $\cN=2$ SCFT $\cT$ one can canonically associate a two-dimensional non-unitary vertex operator algebra (VOA) $\cV[\cT]$ \cite{Beem:2013sza} which arises as a cohomological reduction of the full OPE algebra of the four-dimensional theory. The Schur index of the SCFT is identified up to an overall Casimir-type factor with the VOA's vacuum character, and the structure of $\cV[\cT]$ is deeply connected with the physics of the Higgs branch which conjecturally~\cite{Beem:2017ooy} arises as the associated variety \cite{Arakawa:2010ni} to the VOA (see also \cite{Song:2017oew}). Since the Higgs branch is a symplectic singularity \cite{Hitchin:1986ea,beauville2000symplectic} with finitely many leaves, the VOAs that arise from the cohomological reduction of $\cN=2$ SCFTs would be of a special type known as ``quasi-lisse", a property which as alluded to in the introduction ensures that their vacuum characters $\chi^{}_0$ satisfy a monic modular linear differential equation (MLDE) of finite order~\cite{Arakawa:2016hkg}. The 4d/2d dictionary is more explicitly:
\beq
Z_{\text{Schur}}(q) :=q^{\mathrm{c}/2}\cI(q)= q^{\mathrm{c}/2}{\rm STr}_{\cH}(q^{\Delta-R}) = {\rm STr}_{\cV}(q^{L_0-\mathrm{c}_{2d}/24}) =\chi_0(\tau),
\eeq
with the super-traces $\mathrm{STr}$ referring to traces with $(-1)^F$ insertion, where $F$ is the fermion number in 4d (on $\mathcal{H}$) and the $\mathbb{Z}_2$ grading in 2d (on $\mathcal{V}$). This in turn implies that the Schur index $\cI(q)$, multiplied by the Casimir-type pre-factor $q^{\mathrm{c}/2}$, has interesting modular properties~\cite{Beem:2017ooy}, namely it should transform as a component of a vector-valued modular form of weight zero.\footnote{The question arises whether the other components are also naturally identified with physically interesting quantities. While this has not been fully understood in general, many non-vacuum characters are known to correspond to the Schur index decorated with surface defects orthogonal to the VOA plane \cite{Cordova:2016uwk,Cordova:2017ohl,Cordova:2017mhb,Bianchi:2019sxz, Pan:2021ulr, Pan:2021mrw}.} As we will see, there may be logarithms appearing in the $q$-expansion of the other entries of the vector-valued modular form and therefore more precisely we have a \emph{logarithmic vector-valued modular form}, as defined in \cite{knopp2011logarithmic}. 

Actually, the modular linear differential operator (MLDO) defining the equation satisfied by $\chi^{}_0$ can be modular or twisted modular. In the modular case, the differential equation is invariant under the entire $\mathrm{PSL}(2,\Z)$ group, while in the twisted modular case only under an index-two congruence subgroup $\G^0(2)$. The nature of the MLDE depends on whether half-integer powers appear in the $q$-expansion of $\cI(q)$. If all powers are integers the MLDE is invariant under the full modular group, while it is twisted modular otherwise. From the two dimensional perspective, twisted modular differential operators arise in the presence of operators with half-integer conformal weight in the VOA.\footnote{In the 2d expression for the Schur index $\mathrm{STr}_\mathcal{V}$ really means $\mathrm{Tr}_\mathrm{NS}(-1)^F.$ The modular $T$ transformation $\tau\to\tau+1$ therefore renders the boundary condition along the ``time'' circle anti-periodic for the half-integer modes: $\mathrm{Tr}_\mathrm{NS}(-1)^F\xrightarrow{T} \mathrm{Tr}_\mathrm{NS}$. This is why the Schur index may not be $T$-invariant in presence of half-integer modes. Note that the alternatively defined 2d characters $\mathrm{Tr}_\mathrm{R}(-1)^F$ (corresponding to a decoration of the Schur index with monodromy surface defects in 4d) are $T$-invariant---hence fully $\mathrm{PSL}(2,\Z)$-modular---regardless of presence or absence of half-integer modes since $\mathrm{Tr}_\mathrm{R}(-1)^F\xrightarrow{T}\mathrm{Tr}_\mathrm{R}(-1)^F$. See \cite{Dedushenko:2019yiw} for more details.} We remark, however, that even if there are operators of half-integer weight, it may be the case that cancellations take place in the super-trace, so that the $q$-expansion of $\chi^{}_0$ ends up containing only integer powers implying an untwisted modular differential equation, as noticed in \cite{Beem:2017ooy}.



Let us now discuss the high-temperature expansion (around $q\to1$). In the previous subsection we explained how to obtain the expansion from the integral expression of the index of Lagrangian theories. If it is a general property of Schur indices that they transform as part of a vector-valued modular form (comprising the solutions of the appropriate MLDE) we have an alternative way of analyzing the $q\to1$ expansion which applies to non-Lagrangian $\cN=2$ SCFTs as well  \cite{Beem:2017ooy}. 

We first focus on the modular (as opposed to twisted modular) case. From the $q$-expansion of the Schur index one can algorithmically determine the MLDE that it satisfies by constructing a holomorphic monic modular linear differential operator (see Appendix \ref{App:appendix-modular} for the precise definition and for more details on MLDEs and vector-valued modular forms) with unspecified coefficients and looking systematically for possible solutions. The algorithmic search starts with the smallest possible order of the operator and continues by increasing the order until one finds the first MLDO, of some order $n$, which annihilates the rescaled Schur index,
\begin{equation}
\label{eq:mlde-schur-index}
    \cD^{(n)} Z_{\text{Schur}}(q) = \cD^{(n)} \chi_0(\tau) = 0,
\end{equation}
where by annihilating the Schur index we mean annihilating all the terms up to an appropriately large power (which depends on the order of the MLDE) in the $q$-expansion. 

An MLDE defined by a holomorphic monic MLDO is a Fuchsian equation, with $q=0$ being the only regular singularity inside the unit disk (see \cite{mason-vvmf-mlde} and Appendix~\ref{App:appendix-modular}). The $n$ linearly independent solutions to such an MLDE can be found as power series in $q$ by the usual Frobenius method by first finding the $n$ solutions of the associated indicial equation. We relegate the precise details, which can be found in standard textbooks on differential equations (see \emph{e.g.} \cite{ince1949ordinary}) to Appendix~\ref{App:appendix-modular}. When the solutions $\alpha_i$, $i=0,\dots,n-1$ of the indicial  equation do not differ by an integer, the $n$ solutions to the MLDE will have the form
\begin{equation}
	\chi_i(\tau) = q^{\alpha_i}(a_{i,0} +a_{i,1} q + a_{i,2} q^2+\dots),
	\label{eq:q-expansion-chi-no-log}
\end{equation}
with $a_{i,0}\neq 0$, and where the first solution coincides with the vacuum character $\chi_0(\tau)$, with $\alpha_0 = \mathrm{c}/2$.

When the roots of the indicial equation associated to the MLDE differ by an integer, some solutions might contain powers of $\log q$. For instance, if two of the indicial roots are $\alpha_i$ and $\alpha_i+l$, with $l\in\mathbb{N}_0$, the solution corresponding to $\alpha_i$ will take the form 
\begin{equation}
	\chi_i(\tau) = q^{\alpha_i}(a_{i,0} +a_{i,1} q + a_{i,2} q^2+\dots) + q^{\alpha_i}\log q(b_{i,l}q^l +b_{i,l+1} q^{l+1} +\dots),
	\label{eq:q-expansion-chi-log}
\end{equation}
while the solution corresponding to $\alpha_i+l$ will be a normal $q$-series without logs. If there are more indicial roots differing by integers there will be higher powers of logs and so on, as explained in Appendix~\ref{App:appendix-modular}. 

The vacuum character, $\chi_0(\tau)$, together with the other $n-1$ linearly independent solutions to the MLDE \eqref{eq:mlde-schur-index}, $\chi_{1}(\tau),\dots,\chi_{n-1}(\tau)$, and a multiplier system $\rho$ (to be determined from the vector $\chi(\tau)$), form a weight zero \cite{Beem:2017ooy} weakly holomorphic vector-valued modular form. If some solutions $\chi_i$ contain logarithmic terms, the solutions comprise a logarithmic vvmf \cite{knopp2011logarithmic}. A (logarithmic) vector-valued modular form of weight zero and rank $n$, $\chi(\tau) =(\chi_0(\tau),\dots,\chi_{n-1}(\tau)): \mathbb{H}\to \mathbb{C}^n$, transforms under modular transformations as
\begin{equation}
    \chi\left(\frac{a\tau+b}{c\tau+d} \right) = \rho \begin{pmatrix} a & b \\ c & d
    \end{pmatrix}
  \chi(\tau), \qquad \text{ for } \begin{pmatrix} a & b \\ c & d
    \end{pmatrix} \in \mathrm{PSL}(2,\mathbb{Z}).
  \label{vvmf-transformation1st}
\end{equation}
To simplify the exposition, in this work we will refer to the vector $\chi$ as a vector-valued modular form, although the precise statement should include the adjective \textit{weakly holomorphic}.

 The multiplier system associated to $\chi = (\chi_0,\dots,\chi_{n-1})$ is an $n$-dimensional representation of the modular group  $\mathrm{PSL}(2,\mathbb{Z})$, $\rho: \mathrm{PSL}(2,\mathbb{Z}) \to \mathrm{GL}(n,\mathbb{C})$ which can be obtained demanding that it satisfies \eqref{vvmf-transformation1st} for every $\mathrm{PSL}(2,\mathbb{Z})$ matrix and for any $\tau\in\mathbb{H}$ from the explicit expressions of $\chi_i(\tau)$. 
 
More precisely, to obtain the multiplier system for any $\mathrm{PSL}(2,\Z)$ matrix it is enough to find its form for the two generators $S$ and $T$, since $\rho$ is a group homomorphism and any $\mathrm{PSL}(2,\mathbb{Z})$ matrix can be written as a finite product of $S$ and $T$ matrices. The form of the matrix
\begin{equation}
    \rho(T) \qquad \text{ where }\qquad 
    T = \begin{pmatrix}
        1 & 1 \\ 0 & 1
    \end{pmatrix}
\end{equation}
can be deduced from the $q$ expansions of the kind in \eqref{eq:q-expansion-chi-no-log} and \eqref{eq:q-expansion-chi-log}. In the case with no logarithms, $\rho(T)$ is a diagonal matrix, the diagonal entries being phases given by the roots of the indicial polynomial. In the logarithmic case, the matrix is not  diagonalizable \cite{knopp2011logarithmic} since the $T$ transformation mixes the components with roots that differ by an integer. The $\rho(S)$ matrix cannot be read directly from the vector $\chi(\tau)$ but its form can be constrained by demanding $\rho$ to be a homomorphism (\emph{e.g.}, from the relations that $S$ and $ST$ satisfy) and then one can numerically fix the remaining uncertainties using the vector $\chi(\tau)$ (see the derivation of \eqref{eq:Stransformed-sqcd} for an example with a more detailed explanation of the procedure). 

A weakly holomorphic vvmf can have exponential growth at the cusps, meaning that the $q$-expansions of the components can have a finite number of negative powers. The terms with negative powers (assuming that the matrix $\rho(S)$ mixes all components) control the asymptotic growth of $\chi_0(\tau)$ in the limit $\tau\to 0$ as we explain next.

The special case of \eqref{vvmf-transformation1st} for the $S$ transformation reads
\beq\label{Stransform}
Z_{\text{Schur}}(q) = \chi_0(\tau) = \sum_{i=0}^{n-1} \rho(S)_{0i}\chi_i(-1/\tau), \qquad S = \begin{pmatrix}
    0 & -1\\ 1 & 0
\end{pmatrix}
\eeq
where the matrix elements $\rho(S)_{0i}$ are complex numbers and the $\chi_i(\tau)$ are the components of the vector-valued modular form. The components, in the non-logarithmic cases, can be expanded by power series in $\tilde q$
\beq
\chi_i(-1/\tau)=\tilde q^{h_i}(a_{i,0}+\dots),
\eeq
where the $\dots$ are subleading as $\tilde{q}\to 0$. More generally, there may be solutions that are logarithmic at leading order
\beq
\chi_i(-1/\tau)=\tilde q^{h_i} ((\log \tilde q)^k a_{i,0}+\dots),
\eeq
for some $k \in \N$. Importantly, the leading factors of the non-vacuum characters $\chi_i(\tau)$ contain an energy $h_i$ ($=\,$conformal dimension$\,-\mathrm{c}_{2d}/24$). Note that since the VOAs associated to four dimensional SCFTs are non-unitary, the energies $h_i$ can be less than the vacuum value 
\begin{equation}
    h_0=-\mathrm{c}_{2d} /24=\mathrm{c} /2
\end{equation}
The value of $h_i$ is easily obtained from the solutions $\alpha_i$ of the indicial equation
\begin{equation}
    \alpha_i =h_i.\label{eq:alphaVSh}
\end{equation}

For the twisted case in general the MLDOs do not transform covariantly under all modular transformations but only under $\G^0(2)$ transformations. The matrix $T$ does not belong to $\G^0(2)$ but the matrix $T^2$ does, and therefore the component functions can be expanded in powers of $q^{1/2}$ when $\rho$ is diagonalizable (in the non-diagonalizable case, in powers of $q^{1/2}$ times $(\log q)^k$). The matrix $S$ also does not belong to $\G^0(2)$, so one cannot use equation \eqref{Stransform} to study the $\tau\to 0$ asymptotics. In this case, the asymptotics are given in terms of the solutions of the \textit{conjugate} equation, as follows. 

Let us denote the MLDO by $\mathcal{D}^{}_q$. While in the untwisted case the coefficients of the differentials appearing in $\mathcal{D}^{}_q$ involve  Eisenstein series and thus the $S$-transform of $\mathcal{D}^{}_q$ becomes proportional to itself, in the twisted case $\mathcal{D}^{}_q$ contains the modular forms $\Theta_{r,s}(\tau)$ (see Appendix~\ref{App:appendix-modular}) of the congruence subgroup $\G^0(2)$ as the coefficients of differentials and thus the $S$-transform of $\mathcal{D}^{}_q$ becomes a different operator, that we call the conjugate MLDO and denote by $\tilde{\mathcal{D}}^{}_q$.

More specifically, the modular forms $\Theta_{r,s}(\tau)$ form a basis for the space of all modular forms of $\G^0(2)$ with weight $2r+2s$. Under $S$-transformations, these functions are mapped to the basis elements of the conjugate group $\G_0(2)$,
\begin{equation}
     \Theta_{r,s}\left(-\frac{1}{\tau}\right) = \tau^{2r+2s} \widetilde{\Theta}_{r,s}(\tau).
\end{equation}
Hence, the $\tau\to 0$ behaviour is controlled by the solutions to the MLDE obtained by substituting the basis elements $\Theta_{r,s}(\tau)$ by $\widetilde{\Theta}_{r,s}(\tau)$ and keeping the same coefficients. This is what we call the conjugate MLDE. Thus, in the twisted MLDE case, the $\tau\to 0$ growth is captured by the solutions to the conjugate MLDE, $\tilde{\chi}_i(\tau)$, 
\beq\label{eq:Stransform-twisted}
Z_{\text{Schur}}(q) = \chi_0(\tau) = \sum_{i=0}^{n-1} \rho(S)_{0i}\tilde{\chi}_i(-1/\tau).
\eeq
The components $\tilde{\chi}_i(\tau)$ in this case have expansions
\begin{equation}
	\tilde{\chi}_i(\tau) = q^{\tilde{\alpha}_i}(a_{i,0} +a_{i,1/2} q^{1/2} + a_{i,1} q^1+\dots) + q^{\tilde{\alpha}_i}\log q(b_{i,0} +b_{i,1/2} q^{1/2} + b_{i,1} q^1+\dots)+\dots,
	\label{eq:q-expansion-chi-log-twisted}
\end{equation}
with $a_{i,0}\neq 0$, and with $\tilde{\alpha}_i$  the solutions to the conjugate indicial equation (the indicial equation associated to the conjugate MLDE). These we can write analogously to \eqref{eq:alphaVSh} as
\begin{equation}
    \tilde{\alpha}_i =\tilde{h}_i.
\end{equation}

To summarize, regardless of whether the MLDE is untwisted or twisted, it implies the following structure for the high-temperature (or $\tilde{q}$-) expansion of the Schur index
\begin{equation}
    q^{\mathrm{c}/2}\mathcal{I}(q)=\sum_{i=0}^{n-1}\tilde{q}^{\,\alpha_i}\left(P^{(0)}_{d_i-1}(\log\tilde{q})+\tilde{q}^{1/2}P^{(1/2)}_{d_i-1}(\log\tilde{q})+\tilde{q}\,P^{(1)}_{d_i-1}(\log\tilde{q})
+\cdots\right),\label{eq:MLDEstructuralExp}
\end{equation}
where in the untwisted case $\alpha_i$ are the roots of the indicial equation and $P^{(\text{odd}/2)}_{d_i-1}(\log\tilde q)=0,$ while in the twisted case $\alpha_i$ are the roots of the conjugate indicial equation and $P^{(\text{odd}/2)}_{d_i-1}(\log\tilde q)$ can be nonzero. In either case $P^{(j)}_{d_i-1}(\log\tilde q)$ is a polynomial of order $d_i-1$ in $\log\tilde q$, where $d_i$ is the degeneracy of the root $\alpha_i$ (counting also all the roots that are less than $\alpha_i$ by an integer).

\subsubsection*{Comparison of the structural results}

The structural result \eqref{eq:MLDEstructuralExp} from the MLDE perspective should be compared with the one in \eqref{eq:tauTo0FullAllExpSmall} found from the matrix-integral approach. Note that their compatibility requires that the infinitely-many exponents $h_{\ast_n}$ in \eqref{eq:tauTo0FullAllExpSmall} all be expressible as $\alpha_i$ plus half-integers for finitely-many $\alpha_i.$ Fortunately, this is guaranteed by Remark~2 above.

On the other hand, the fact implied by the matrix-integral result \eqref{eq:tauTo0FullAllExpSmall} that there should naturally be a term with exponent $\tilde{q}^{2(\mathrm{a}-\mathrm{c})}$ in the expansion, as emphasized in Remark~1 above, appears impossible to see from the MLDE perspective \eqref{eq:MLDEstructuralExp}. In the opposite direction, the fact implied by the MLDE that there should naturally be an $h_{\ast_n}=\alpha_0=\mathrm{c}/2$ appears impossible to see from the matrix-integral perspective.


We mention in passing another set of works on the quasi-modular expressions for the Schur index \cite{Beem:2021zvt, Pan:2021mrw, Huang:2022bry}. In \cite{Beem:2021zvt} it was shown that the Schur index for a family of theories can be expressed in terms of mixed-weight quasi-modular forms. This was shown to be consistent with the MLDE perspective since a quasi-modular form of weight $k$ and depth $p$ can be seen as a component of a logarithmic vvmf under the weight $k-p$ action of the modular group. Under this action,  the $q$ expansion of a quasi-modular form of weight $k$ and depth $p$ has the form of a polynomial in $\log q$ of degree $\leq p$. From our structural results derived from the matrix integral perspective \eqref{eq:tauTo0FullAllExpSmall}, we can predict a rough bound on the possible depth of the quasi-modular forms appearing in the Schur index. Namely, the depth $p$ of these quasi-modular forms must satisfy $p \leq \mathrm{rank}$.

\subsection{Examples}\label{subsec:ExSec2}

\subsubsection*{The SU($2$) $\mathcal{N}=4$ theory} 

This theory has $\mathrm{a}=\mathrm{c}=3/4.$

The Rains function of the $\mathcal{N}=4$ theory identically vanishes. So $L^{\mathrm{min}}=0$ and $\mathrm{dim}\mathfrak{h}_\text{qu}=1$. The $\tau\to0$ asymptotic follows from \eqref{eq:tauTo0FullAllExpSmall} to be of the form
\begin{equation}
    e^{i\pi\tau\frac{3}{4}}\ \mathcal{I}(q)\simeq
   \frac{A}{-i\tau}+B,\label{eq:N=4tauTo0Full}
\end{equation}
up to exponentially suppressed corrections, and with $A,B$ to be determined. (Using the Fermi gas results \cite{Bourdier:2015wda} it was found in \cite{ArabiArdehali:2015ybk} that $A=1/4,$ $B=-1/2\pi$.) Note that since the Rains function $L_h$ vanishes in this case it does not yield any exponentially small corrections to \eqref{eq:N=4tauTo0Full}.

The higher Rains functions are linear combinations of non-negative integer multiples of the generators $\{2x\},1-\{2x\},1$, all of which lead to critical sets with $L_\ast$ a positive integer and $\mathrm{dim}\mathfrak{h}^\ast=0$ or $1$. In particular, the first exponentially suppressed correction to \eqref{eq:N=4tauTo0Full} modifies it to
\begin{equation}
    e^{i\pi\tau\frac{3}{4}}\ \mathcal{I}(q)=
   \frac{A}{-i\tau}+B\, +e^{-\frac{2\pi i}{\tau}}\cdot\bigg(\frac{C}{-i\tau}+D\bigg)+\mathcal{O}\big(e^{-\frac{4\pi i}{\tau}}\big),\label{eq:N=4tauTo0FirstExpSmall}
\end{equation}
with $C,D$ to be determined.

This much we could say with our ``semi-qualitative''  analysis based on \eqref{eq:tauTo0FullAllExpSmall}. However, our algorithmic procedure that led to \eqref{eq:tauTo0FullAllExpSmall} can go further and determine the coefficients $A,B,C,D$ as well. We illustrate this ``fully quantitative'' analysis in the present case, but for the remaining examples below we content ourselves with the semi-qualitative treatment.

Our fully quantitative analysis starts with the matrix integral \eqref{eq:indexStransformed}, which for the SU($2$) $\mathcal{N}=4$ theory reads
\begin{equation}
    \begin{split}
  e^{i\pi \tau \mathrm{c}}\,\mathcal{I}(q)&= \frac{1}{2}\, \bigg(\prod_{k=1}^\infty \frac{(1-\tilde{q}^k)^{2}}{(1+\tilde{q}^k)^{2}} \bigg)\cdot  \frac{1}{-i\tau}\int_{-1/2}^{1/2} \frac{\mathrm{d}x}{2}\,  \ \prod_{l=1}^\infty\, \frac{ (1-\tilde{q}^{l-\{2x\}})^2\, (1-\tilde{q}^{l-1+\{2x\}})^2}{ (1+\tilde{q}^{l-\{2x\}})^2(1+\tilde{q}^{l-1+\{2x\}})^2}.
    \end{split}\label{eq:N=4indexStransformed}
\end{equation}

The integration domain can be decomposed into an outer patch consisting of two copies of $(\frac{\epsilon}{2},\frac{1}{2}-\frac{\epsilon}{2})$, and two inner patches each consisting of two copies of $(0,\frac{\epsilon}{2})$,  thanks to the Weyl and center symmetry. The inner patch contribution (from the four copies of $(0,\frac{\epsilon}{2})$) is
\begin{equation}
\begin{split}
      e^{i\pi \tau \mathrm{c}}\mathcal{I}_{\mathrm{in}}(q)&=\frac{1}{2}(1-4\tilde{q})\cdot\frac{1}{-i\tau}\times 4\times \int^{\frac{\epsilon}{2}}_{0}\frac{\mathrm{d}x}{2}\ \frac{(1-2\tilde{q}^{2x}+\tilde{q}^{4x})(1-4\tilde{q}^{1-2x}-4\tilde{q}^{1+2x})}{(1+\tilde{q}^{2x})^2} +o(\tilde{q})\\
      &=\frac{1-4\tilde{q}}{2\pi}\big[-4\tilde{q}I_1(\Lambda)+(1+8\tilde{q})I_2(\Lambda)-(2+8\tilde{q})I_3(\Lambda)+(1+8\tilde{q})I_4(\Lambda)-4\tilde{q} I_5(\Lambda)\big]\\
      &\ \ \ \ +o(\tilde{q}),
      \end{split}
\end{equation}
where
\begin{equation}
    I_j(\Lambda):=\int_0^{\frac{\Lambda}{2}} \frac{e^{2(2-j)x}}{(1+e^{-2x})^2}.
\end{equation}
Integrating the RGE of $I_j(\Lambda)$ and determining the integration constants via the subtraction method\footnote{To obtain the result for $I_1$ we note that its integrand approaches $e^{2x}-2$ as $x\to\infty$. In the \emph{subtraction method} we write
\begin{equation}
    I_1(\Lambda)=\int_0^{\frac{\Lambda}{2}}\mathrm{d}x\ \left(\frac{e^{2x}}{(1+e^{-2x})^2}-e^{2x}+2\right)+ \int_0^{\frac{\Lambda}{2}}\mathrm{d}x\ (e^{2x}-2).
\end{equation}
The first integral now can be written as $\int_0^\infty-\int_{{\Lambda}/{2}}^\infty$; the part $\int_0^\infty$ gives only a constant $1/4+\log2$ thanks to the convergence achieved by our subtraction, while the part $\int_{{\Lambda}/{2}}^\infty$ is $O(e^{-\Lambda})$ since its integrand can be safely Taylor-expanded. The second integral has a constant piece $-1/2$. Putting them together, the integration constant of the RGE is found to be $-1/4+\log2.$ Similar analysis yields the constant pieces of $I_{2,3,4,5}(\Lambda)$. See Section~7.2 of \cite{Wilson:1973jj} for Wilson's use of this method in a non-perturbative derivation of near-critical scaling laws. \label{ftnt:subtractionMethod}} we find
\begin{equation}
    \begin{split}
    I_1(\Lambda)&=e^{\Lambda}/2-\Lambda-\frac{1}{4}+\log2,\\
        I_2(\Lambda)&={\frac{\Lambda}{2}}-\frac{1}{4}-\log2/2,\\
        I_3(\Lambda)&=\frac{1}{4},\\
        I_4(\Lambda)&=-\frac{1}{4}+\log2/2,\\
        I_5(\Lambda)&=\frac{3}{4}-\log2,\\
        \end{split}
\end{equation}
up to $O(e^{-\Lambda})$, and hence
\begin{equation}
    e^{i\pi \tau \mathrm{c}}\mathcal{I}_{\mathrm{in}}(q)=\frac{1}{2\pi}[(-1+{\frac{\Lambda}{2}})+4\tilde{q}(-1+\Lambda)]+o(\tilde{q}).\label{eq:N=4fullQuantIntApproach}
\end{equation}
The outer patch analysis is straightforward, and the addition of that contribution turns out to complete $2\epsilon$ to $1$ (hence $2\Lambda$ to $\frac{2\pi i}{\tau}=-\log\tilde{q}$), so we get
\begin{equation}
    e^{i\pi \tau \mathrm{c}}\mathcal{I}(q)=\frac{1}{2\pi}[(-1-\frac{1}{4}\log\tilde{q})+4\tilde{q}(-1-\frac{1}{2}\log\tilde{q})]+o(\tilde{q}).\label{eq:SU(2)N=4RGderivedAsy}
\end{equation}
Note that the first two terms on the RHS are in agreement with the result of \cite{ArabiArdehali:2015ybk}, and the pattern of logarithms compatible with that found in \cite{Dedushenko:2019yiw}.\footnote{The numbers in \eqref{eq:SU(2)N=4RGderivedAsy} actually do not exactly match with those reported in the 1st version of \cite{Dedushenko:2019yiw}. We are grateful to M.~Dedushenko for confirming that the mismatch is due to the fact that the inner-patch contributions were not properly taken into account in that work.}\\

We now adopt the MLDE approach. The modular differential equation associated to the Schur index of this theory is found to be of second order twisted, and it is given by the following MLDO
\begin{align}
\cD_{SU(2)\ \cN=4}&=D_q^{(2)}-\frac1{12}\Theta_{0,1} D_q^{(1)} + \frac3{64} (\Theta_{1,1}-\Theta_{0,2}).
\label{MLDE-su2n4}
\end{align}
See Appendix~\ref{App:appendix-modular} for the definition of $D_q^{(k)}$ and $\Theta_{j,k}.$ The associated indicial equation is
\begin{equation}
    \alpha^2-\frac{\alpha}{4}-\frac{3}{64}=0,
    \label{indicial-su2n4}
\end{equation}
with solutions $\alpha = -1/8,3/8$. Since the MLDO only transforms properly under $\G^0(2)$ transformations, the $\tau\to 0$ growth is given by the indicial equation of the conjugate MLDE,
\begin{align}
    \tilde{\cD}_{SU(2)\ \cN=4}&=D_q^{(2)}-\frac1{12}\tilde{\Theta}_{0,1} D_q^{(1)} + \frac3{64} (\tilde{\Theta}_{1,1}-\tilde{\Theta}_{0,2}).
    \label{conjugate-MLDE-su2n4}
\end{align}
The conjugate indicial equation is simply
\begin{equation}
    \tilde{\alpha}^2 = 0,
    \label{indicial-conjugate-su2n4}
\end{equation}
which, as explained in Appendix \ref{App:appendix-modular}, the double root implies that the two solutions $\tilde{\chi}^{}_0(\tau), \tilde{\chi}_1(\tau)$ will be a power series in $q$ and a power series times $\log(q)$, respectively. Therefore, the predicted growth near $\tau\to 0$ of the Schur index \eqref{eq:Stransform-twisted} yields
\begin{equation}
   e^{i\pi\tau\frac{3}{4}} \cI(q) = \sum_i \cS_{0i}\tilde{\chi}_i(-1/\tau) =  \frac{A}{-i\tau}+B\, +e^{-\frac{2\pi i}{\tau}}\cdot\bigg(\frac{C}{-i\tau}+D\bigg)+\mathcal{O}\big(e^{-\frac{4\pi i}{\tau}}\big),\label{eq:N=4tauto0FromMLDE}
\end{equation}
which precisely agrees with \eqref{eq:N=4tauTo0FirstExpSmall}.

Our results can be checked against the closed form expression of the Schur index in terms of quasimodular forms of \cite{Beem:2021zvt},
\begin{equation}
    e^{i\pi\tau\frac{3}{4}}\ \mathcal{I}(q) = \frac{\eta(\tau/2)^2}{\eta(\tau)^4}\left( \frac{\E_2(\tau)}{2} +\frac{\Theta_{0,1}(\tau)}{24}\right).
\end{equation}
To obtain the high-temperature expansion we use the $S$-transformed expressions
\begin{equation}
    \Theta_{r,s}(-1/\tau) = \tau^{2r+2s}\tilde{\Theta}_{r,s}(\tau),  \qquad \eta(-1/\tau) = \sqrt{-i\tau} \eta(\tau),
\end{equation}
and the anomalous transformation of $\E_2(\tau)$ given in \eqref{eq:e2anomalous} to rewrite the Schur index: 
\begin{equation}
    e^{i\pi\tau\frac{3}{4}}\ \mathcal{I}(q) = \frac{\eta(-2/\tau)^2}{\eta(-1/\tau)^4}\frac{1}{2\pi}\left( \left( {\E_2(-1/\tau)} +\frac{\tilde{\Theta}_{0,1}(-1/\tau)}{12}\right)\log \tilde{q}-1\right),
\end{equation}
which we can expand to get the precise numerical coefficients
\begin{equation}
\begin{split}
     e^{i\pi\tau\frac{3}{4}}\ \mathcal{I}(q)= & 
     -\frac{1}{8\pi}\left(
     {4+\log \tilde{q}}
      +{ (16+12 \log \tilde{q})\tilde{q}}
      +o(\tilde{q})\right)
\end{split}
\label{eq:N=4tauto0FromQuasi}
\end{equation}
with the same form as in \eqref{eq:N=4tauTo0FirstExpSmall} and \eqref{eq:N=4tauto0FromMLDE}, and the same coefficients as in \eqref{eq:N=4fullQuantIntApproach}.

\subsubsection*{SU(2) SQCD}

This theory has $\mathrm{a}-\mathrm{c}=-5/24$ and $\mathrm{c}=7/6.$

\begin{figure}[h]
\centering
    \includegraphics[scale=.45]{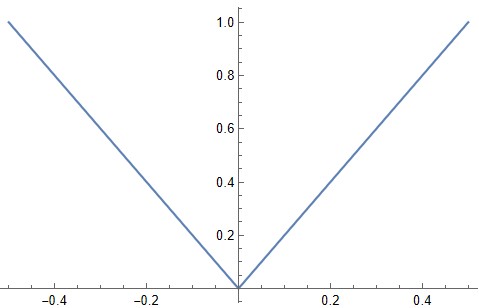}
\caption{Rains function $L(x)$ of SU(2) SQCD with $N_f=4$. 
\label{fig:su2sqcd}}
\end{figure}

The Rains function looks as in Fig~\ref{fig:su2sqcd}. So we have $L^{\mathrm{min}}=0$ and $\mathrm{dim}\mathfrak{h}_\text{qu}=0$. Therefore the small-$\tau$ asymptotic looks like
\begin{equation}
    e^{i\pi{\tau}\frac{7}{6}}\ \mathcal{I}(q)\simeq
   e^{\frac{5i\pi}{6{\tau}}}\cdot A,\label{eq:SQCDtauTo0Full}
\end{equation}
up to exponentially suppressed corrections. The other extremum of the Rains function $L(x)$ in this case is at $x=\pm 1/2$ (with $\mathrm{dim}\mathfrak{h}^\ast=0$) and has the value $L_\ast=1$. Hence \eqref{eq:tauTo0FullAllExpSmall} implies that \eqref{eq:SQCDtauTo0Full} receives a correction as
\begin{equation}
    e^{i\pi{\tau}\frac{7}{6}}\ \mathcal{I}(q)=
   e^{\frac{5i\pi}{6{\tau}}}\cdot A+e^{-\frac{7\pi i}{6\tau}}\cdot C+\cdots.\label{eq:SQCDtauTo0ExpSmallSome}
\end{equation}

\begin{figure}[h]
\centering
    \includegraphics[scale=.47]{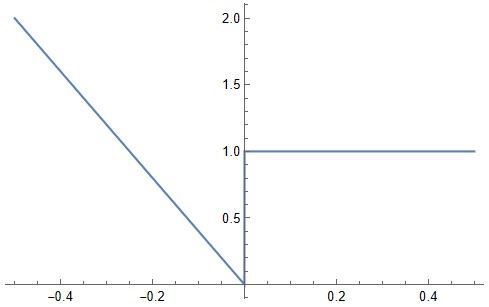}
\caption{Higher Rains function $L^{(2)}(x)$ of SU(2) SQCD with $N_f=4$ given by the Rains function plus the generator $1-\{\alpha_+\cdot\boldsymbol{x}\}=1-\{2x\}$. 
\label{fig:su2sqcdHigher}}
\end{figure}

\begin{figure}[h]
\centering
    \includegraphics[scale=.47]{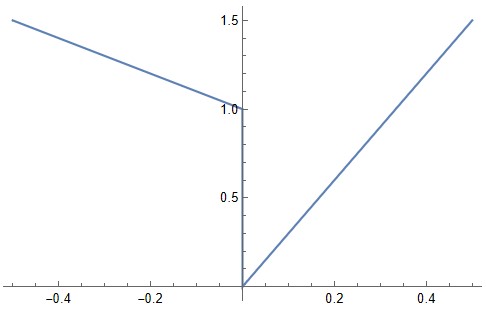}
\caption{Higher Rains function $L^{(3)}(x)$ of SU(2) SQCD with $N_f=4$ given by the Rains function plus the generator $\{\rho^\chi_+\cdot\boldsymbol{x}\}=\{x\}$. 
\label{fig:su2sqcdHigher3}}
\end{figure}

The higher Rains functions are $L(x)$ plus linear combinations with non-negative integer coefficients of the generators $\{x\},1-\{x\},\{2x\},1-\{2x\},1$, all of which lead to vertex values $L_\ast$ larger than or equal to $L_\ast=1$. There is in particular a higher Rains function with $L_\ast=1$ and $\mathrm{dim}\mathfrak{h}^\ast=1$, as in Figure~\ref{fig:su2sqcdHigher}. This will 
modify \eqref{eq:SQCDtauTo0ExpSmallSome} to
\begin{equation}
    e^{i\pi{\tau}\frac{7}{6}}\ \mathcal{I}(q)=
   e^{\frac{5i\pi}{6{\tau}}}\cdot A+e^{-\frac{7\pi i}{6\tau}}\cdot \left(\tilde{B}\, \log\tilde{q}+\tilde{C}\right)+\mathcal{O}(e^{-\frac{13\pi i}{6\tau}}),\label{eq:SQCDtauTo0ExpSmallFirstSub}
\end{equation}
with the error found from examination of other higher Rains functions (in particular $L^{(3)}(x)=L(x)+\{x\}$ which has a vertex value $3/2$ with $\mathrm{dim}\mathfrak{h}^\ast=0$ as in Figure~\ref{fig:su2sqcdHigher3}). The result \eqref{eq:SQCDtauTo0ExpSmallFirstSub}, and especially the pattern of logarithms in it, is compatible with Eq.~(5.15) of \cite{Dedushenko:2019yiw}. 

In fact the result of \cite{Dedushenko:2019yiw} implies that the error term in \eqref{eq:SQCDtauTo0ExpSmallFirstSub} should be not just $\mathcal{O}(e^{-\frac{13\pi i}{6\tau}}),$ but  $\mathcal{O}(e^{-\frac{19\pi i}{6\tau}}).$ To see this in our approach we need a more careful analysis of potential cancellations. This will be discussed briefly in Section~\ref{sec:Discussion}.

Now we adopt the MLDE approach. In this case the modular differential equation is found to be of untwisted second order, and it is given by the following MLDO
\begin{align}
\cD^{}_{SU(2)\ N_f=4}&=D_q^{(2)}-
175\, \mathbb{E}_4.
\label{su2qcd-MLDE}
\end{align}
The indicial equation arising from the MLDE is 
\begin{equation}
    \alpha^2-\frac{\alpha}{6}-\frac{35}{144}=0,
\end{equation}
with solutions $\alpha_0 = \frac{7}{12}$, $ \alpha_1 = -\frac{5}{12}$. Since $\alpha_0 = \alpha_1 +1$, the solution associated to the root $\alpha_1$ can contain logarithmic terms. Using the Fuchsian ansatz for the solutions
\begin{equation}
    \begin{cases}
        \chi^{}_0(\tau) = q^{\alpha_0}h_0(q) \\
        \chi_1(\tau) = q^{\alpha_1}h_1(q)+A \log(q)\chi^{}_0(\tau),
    \end{cases}
\end{equation}
we normalize the vacuum character to $\chi^{}_0(\tau) =   e^{i\pi{\tau}\frac{7}{6}}\ \mathcal{I}(q)$, and obtain
\begin{equation}
\begin{split}
        \chi_1(\tau) = & q^{-\frac{5}{12}} \left(1 - 2375 q^2 - 38000 q^3 - 352250 q^4 - 2391248 q^5 - 
   13290625 q^6-\dots\right) \\
 &+60 \log(q)\chi_1(\tau)
\end{split}
\end{equation}
for the module character, where the constant $A$ has been fixed to be the smallest number such that the character has an integral expansion. From these solutions, assuming that the $S$ transformation mixes the two solutions, the growth near $\tau\to 0$ will have the following schematic form
\begin{equation} \label{eq:asymptotic-lmde-su2sqcd}
   e^{i\pi{\tau}\frac{7}{6}}\ \mathcal{I}(q) =	\chi^{}_0(\tau) = A e^{\frac{5}{12}\frac{2\pi i}{\tau}} + \left( B + C\log \tilde{q} \right)e^{-\frac{7}{12}\frac{2\pi i}{\tau}} + \dots,
\end{equation}
reproducing \eqref{eq:SQCDtauTo0ExpSmallFirstSub}. 

From the MLDE perspective one can obtain more explicit answers. In particular, it is possible to compute the precise coefficients of equation \eqref{eq:asymptotic-lmde-su2sqcd}, as follows. The vector-valued modular form will transform as
\begin{equation}
    \chi\left(\frac{a\tau+b}{c\tau+d} \right) = \rho \begin{pmatrix} a & b \\ c & d
    \end{pmatrix}
  \chi(\tau),
  \label{vvmf-transformation}
\end{equation}
where the multiplier system $\rho$ is a two-dimensional representation of $\mathrm{PSL}(2,\mathbb{Z})$, $\rho: \mathrm{PSL}(2,\mathbb{Z}) \to \mathrm{GL}(2,\mathbb{C})$. From the explicit form of the solutions $\chi(\tau)$, we obtain
\begin{equation}
    \rho(T) = \rho \begin{pmatrix} 1 & 1 \\ 0 & 1
    \end{pmatrix} = e^{2\pi i \frac{7}{12}}\begin{pmatrix}
        1 & 0 \\
        120\pi i  & 1
    \end{pmatrix}.
\end{equation}
Requiring that $\rho$ is a homomorphism, from the relations satisfied by the generators of the $\mathrm{SL}(2,\mathbb{Z})$ group, $S^2= -I_2 = (ST)^3$, and the following property of weight zero multiplier systems, $\rho(\pm I_2) = I_2$, the matrix $\rho(S)$ gets fixed up to one parameter.  Using the explicit knowledge of the vector of solutions, which can be computed to arbitrary orders in the $q$ expansion from the MLDE, one can numerically fix the remaining parameter in the matrix $\rho(S)$, since \eqref{vvmf-transformation} is an equation valid for all $\tau$. One then finds that the $S$ transformation mixes the characters in the following way
\begin{equation}
\begin{split}
      \chi^{}_0(\tau)  = &  \frac{25}{12 \pi }\chi^{}_0(-1/\tau)+\frac{1}{120 \pi }\chi_1(-1/\tau) \\
       = &  \frac{ \tilde{q}^{-5/12}}{120 \pi }\left(
       1+250 \tilde{q}+4625 \tilde{q}^2+44250 \tilde{q}^3+305750 \tilde{q}^4+1703752 \tilde{q}^5 + \dots\right.
      \\
      &\left.+ \log(\tilde{q})(60 \tilde{q} + 1680 \tilde{q}^2 + 19740 \tilde{q}^3 + 157920 \tilde{q}^4 + 982800 \tilde{q}^5 + \dots)
      \right),
\end{split}
\label{eq:Stransformed-sqcd}
\end{equation}
which matches the results of \cite{Dedushenko:2019yiw}.

An alternative check of the coefficients of this expansion comes from the expression of the Schur index in terms of quasimodular forms of \cite{Beem:2021zvt},
\begin{equation}
     e^{i\pi{\tau}\frac{7}{6}}\mathcal{I}(q) = \eta(\tau)^{-10}\big( -12\, \E_2(\tau)\E_4(\tau) + 42\,\E_6(\tau)\big).
\end{equation}
From the anomalous transformation of $\E_2(\tau)$ given in \eqref{eq:e2anomalous} and the modular transformation properties of the other functions one obtains
\begin{equation}
    e^{i\pi{\tau}\frac{7}{6}} \mathcal{I}({q}) = \frac{\eta(-1/\tau)^{-10}}{2\pi}\big( \log(\tilde{q})\big(-12  \E_2(-1/\tau)\E_4(-1/\tau) + 42\E_6(-1/\tau)\big) + 12 \E_4(-1/\tau)\big),
\end{equation}
and expanding it one gets precisely \eqref{eq:Stransformed-sqcd}.

\subsubsection*{USp(4) with half-hyper in 16}
This theory has $24a =58$ and $12c=28$, hence $\mathrm{a}-\mathrm{c}=1/12$.

\begin{figure}[h]
\centering
    \includegraphics[scale=.5]{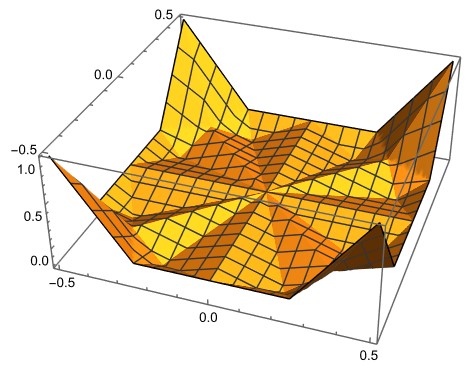}
\caption{Rains function $L(x_1,x_2)$ of the USp(4) theory.
\label{fig:usp4}}
\end{figure}

The Rains function looks as in Fig~\ref{fig:usp4}. So we have $L^{\mathrm{min}}=-1/5$ and $\mathrm{dim}\mathfrak{h}_\text{qu}=0$. Therefore \eqref{eq:tauTo0FullAllExpSmall} implies the leading asymptotic
\begin{equation}
    e^{i\pi \tau \mathrm{c}}\ \mathcal{I}(q)\simeq e^{-\frac{2\pi i}{ \tau}[2(\mathrm{a}-\mathrm{c})-\frac{1}{5}]}\cdot C.\label{eq:USp4tauTo0Full}
\end{equation}
Examination of Figure~\ref{fig:usp4} shows that the other extrema of the Rains function are at the origin (with $L_\ast=0$ and $\mathrm{dim}\mathfrak{h}^\ast=1$) and at $x_1=\pm x_2=1/2$ (with $L_\ast=1$ and $\mathrm{dim}\mathfrak{h}^\ast=0$). Hence \eqref{eq:tauTo0FullAllExpSmall} implies that \eqref{eq:USp4tauTo0Full} is corrected to
\begin{equation}
    e^{i\pi \tau \mathrm{c}}\ \mathcal{I}(q)= e^{-\frac{2\pi i}{ \tau}(2(\mathrm{a}-\mathrm{c})-\frac{1}{5})}\cdot C+e^{-\frac{2\pi i}{ \tau}(2(\mathrm{a}-\mathrm{c}))}\cdot \left(\frac{A}{-i\tau}+B\right) +e^{-\frac{2\pi i}{ \tau}(2(\mathrm{a}-\mathrm{c})+1)}\cdot D +\cdots.\label{eq:USp4tauTo0ExpSmallSome}
\end{equation}

\begin{figure}[h]
\centering
    \includegraphics[scale=.55]{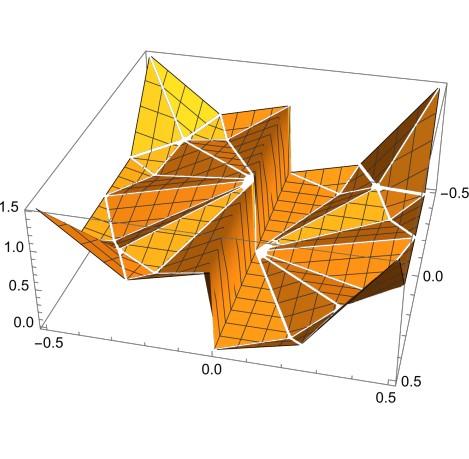}
\caption{Higher Rains function $L^{(2)}(x_1,x_2)$ of the USp(4) theory given by the first Rains function plus the generator $\{\alpha_+\cdot\boldsymbol{x}\} = \{x_2\} $.
\label{fig:usp4-higher}}
\end{figure}

By inspection of the higher Rains function given in Figure \ref{fig:usp4-higher}, we see the existence of degenerate minima with value $L_\ast=0$ and ${\mathrm{dim}\mathfrak{h}^{\ast}}=2$. This will change the degree-one polynomial in $\tau^{-1}$ multiplying $e^{-\frac{2\pi i}{ \tau}(2(\mathrm{a}-\mathrm{c}))}$ in \eqref{eq:USp4tauTo0ExpSmallSome} to a polynomial of degree two,
\begin{equation}
    e^{i\pi \tau \mathrm{c}}\ \mathcal{I}(q)= e^{-\frac{2\pi i}{ \tau}(2(\mathrm{a}-\mathrm{c})-\frac{1}{5})}\cdot C+e^{-\frac{2\pi i}{ \tau}(2(\mathrm{a}-\mathrm{c}))}\cdot \left(\frac{\tilde{A}}{(-i\tau)^2}+\frac{\tilde{B}}{-i\tau}+\tilde{C}\right) +\mathcal{O}\big(e^{-\frac{2\pi i}{ \tau}(2(\mathrm{a}-\mathrm{c})+\frac{1}{5})}\big),\label{eq:USp4tauTo0ExpSmallhigher}
\end{equation}
with the error following from examination of vertex values of $L^{(2)}(x_1,x_2).$

We now adopt the MLDE approach. By direct computation we find that $\cI_{\mathrm{USp}(4)}(q)$ satisfies the following untwisted MLDE of order seven:
\begin{align}
\cD_{\mathrm{USp}(4)}&=D_q^{(7)}-\frac{2344}5\E_4D_q^{(5)}-\frac{30072}5\E_6 D_q^{(4)}+34000\,\E_3^2 D_q^{(3)}-17248\,\E_4\E_6 D_q^{(2)}\\
&\qquad-\left(126720\, \E_4^3+232848\, \E_6^2\right) D_q^{(1)}-1552320 \,\E_4^2\E_6
\end{align}

All in all we expect the MLDE satisfied by this index to be of dimension seven with at least two degenerate eigenvalues. In fact the degenerate eigenvalues are three, one coming from the first piece and two from the second piece.

The indicial equation derived from the MLDE is
\begin{equation}
	\alpha ^7-\frac{7 \alpha ^6}{2}+\frac{421 \alpha ^5}{100}-\frac{11441 \alpha ^4}{5400}+\frac{5257 \alpha ^3}{10800}-\frac{341 \alpha ^2}{7200}+\frac{179 \alpha }{233280}+\frac{77}{777600}=0,
\end{equation}
and has the following solutions, displayed in increasing order and with multiplicity,
\begin{equation}
\alpha = -\frac{1}{30},\, \frac{1}{6},\,\frac{1}{6},\,\frac{1}{6},\,\frac{11}{30}, \,\frac{7}{6},\,\frac{3}{2}.
\end{equation}
The smallest solution reproduces the leading exponential growth, since $2(\mathrm{a}-\mathrm{c})-1/5=-1/30$. The four roots $\frac{1}{6},\,\frac{1}{6},\,\frac{1}{6}, \,\frac{7}{6}$ differ by an integer, and therefore a priori we could get terms $(\log q)^3$ in one of the associated solutions. Therefore a more careful analysis is needed to show that only terms with $(\log q)^2$ appear.

A closed form expression for the Schur index of this theory is conjectured by Shlomo~Razamat.\footnote{We thank him warmly for sharing it with us.} It reads
\beq
\cI_{\mathrm{USp}(4)}(q)=\frac{q^{-7/6}}{5}\left(\frac{\eta(5\tau)}{\eta(\t)}-\eta^4(\tau)\right).\label{eq:ShlomoConj}
\eeq

From this expression we can get the different contributions to the high temperature expansion by using the modular properties of the Dedekind eta function. In particular, one can see directly that the $(\log q)^2$ terms will come from the second summand. More explicitly, from
\begin{equation}
	\eta(-1/\tau) = \sqrt{-i\tau} \eta(\tau),
\end{equation}
we obtain
\beq
\cI_{\mathrm{USp}(4)}(q)=\frac{q^{-7/6}}{5}\left(\frac{1}{\sqrt{5}} 	\frac{\eta(-\frac{1}{5\tau})}{\eta(-\frac{1}{\t})}+\frac{1}{\tau^2}\eta\left(-\frac{1}{\t}\right)^4\right).
\eeq
The expansion in $\tilde{q}$ in the limit $\tau\to 0$ of this expression provides the high temperature expansion with the exact coefficients,
\beq
\begin{split}
    \cI_{\mathrm{USp}(4)}(q) =&\frac{q^{-7/6}}{5}\left(\frac{1}{\sqrt{5}}\left(\tilde{q}^{-1/30}-\tilde{q}^{1/6}-\tilde{q}^{11/30}+2 \tilde{q}^{29/30}-\tilde{q}^{7/6}+3 \tilde{q}^{59/30}+\dots \right)\right.\\
   & \left.+\frac{1}{\tau^2}\left(\tilde{q}^{1/6} -4 \tilde{q}^{7/6}+2 \tilde{q}^{13/6}+8 \tilde{q}^{19/6}+\dots \right)\right).
\end{split}
\eeq
This expression agrees with the fact, coming from the Rains functions analysis, that there are no $\tau^{-3}$ terms in the high temperature expansion. It moreover implies that the terms proportional to $\tau^{-1}$ should vanish. In the integral approach the latter vanishing corresponds to a completely destructive interference phenomenon which our semi-qualitative analysis can not see, but our algorithmic (fully quantitative) approach should capture.

\subsubsection*{SU$(2)^2$ with one $(2, 2)$, two $(2, 1)$s, and two $(1, 2)$s}

This theory has $\mathrm{a}-\mathrm{c}=-1/4$.

\begin{figure}[h]
\centering
    \includegraphics[scale=.45]{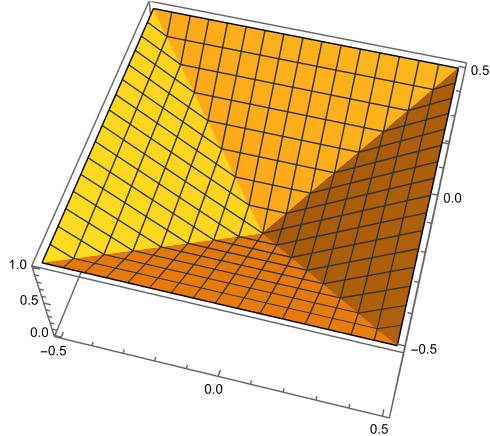}
\caption{Rains function $L(x_1,x_2)$ of the SU(2)$\times$SU(2) theory.
\label{fig:su2xsu2}}
\end{figure}

The Rains function looks as in Fig~\ref{fig:su2xsu2}. We have $L^{\mathrm{min}}=0$ and $\mathrm{dim}\mathfrak{h}_\text{qu}=0$. Therefore
\begin{equation}
    e^{i\pi \tau \mathrm{c}}\  \mathcal {I}(q)\simeq e^{-\frac{2\pi i}{ \tau}(2(\mathrm{a}-\mathrm{c}))}\cdot C.\label{eq:su2xsu2tauTo0Full}
\end{equation}
Examination of Figure~\ref{fig:su2xsu2} shows that the other extremum of the Rains function is at $x_1=x_2= 1/2$ (with $L^\ast=1$ and $\mathrm{dim}\mathfrak{h}^\ast=1$). Hence \eqref{eq:tauTo0FullAllExpSmall} implies that \eqref{eq:su2xsu2tauTo0Full} is corrected to
\begin{equation}
    e^{i\pi \tau \mathrm{c}}\ \mathcal{I}(q)= e^{-\frac{2\pi i}{ \tau}(2(\mathrm{a}-\mathrm{c}))}\cdot C+e^{-\frac{2\pi i}{ \tau}(2(\mathrm{a}-\mathrm{c})+1)}\cdot \left(\frac{A}{-i\tau}+B\right)+\cdots.\label{eq:su2xsu2tauTo0ExpSmallSome}
\end{equation}

\begin{figure}[h]
\centering
    \includegraphics[scale=.45]{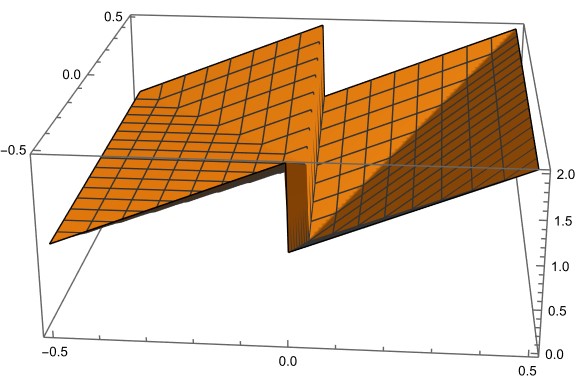}
\caption{Rains function $L^{(2)}(x_1,x_2)=L(x_1,x_2)$ of the SU(2)$\times$SU(2) theory.
\label{fig:su2xsu2Higher}}
\end{figure}

There is a higher Rains function $L^{(2)}(x_1,x_2)=L(x_1,x_2)+\{2x_1\}$, that has a vertex value $L^\ast=1$ with $\mathrm{dim}\mathfrak{h}^\ast=2$, as can be seen in Firgure~\ref{fig:su2xsu2Higher}. This will modify \eqref{eq:su2xsu2tauTo0ExpSmallSome} to
\begin{equation}
    e^{i\pi \tau \mathrm{c}}\ \mathcal{I}(q)= e^{-\frac{2\pi i}{ \tau}(2(\mathrm{a}-\mathrm{c}))}\cdot C+e^{-\frac{2\pi i}{ \tau}(2(\mathrm{a}-\mathrm{c})+1)}\cdot \left(\frac{\tilde{A}}{(-i\tau)^2}+\frac{\tilde{B}}{-i\tau}+\tilde{C}\right)+\mathcal{O}\big(e^{-\frac{2\pi i}{ \tau}(2(\mathrm{a}-\mathrm{c})+3/2)}\big),\label{eq:su2xsu2tauTo0ExpSmallFirstSub}
\end{equation}
with the error again following from examination of vertex values of higher Rains functions.

In this case instead the modular differential equation is found to be of the fourth order:
\begin{align}
\cD_{\mathrm{SU}(2)\times \mathrm{SU}(2)}&=D_q^{(4)}+
\left(\frac{11}{72} \Theta_{1,1}-\frac{11}{36}\Theta_{0,2}\right)D_q^{(2)}-
\left(\frac{5}{216} \Theta_{0,3}+\frac{13}{144}\Theta_{1,2}\right)D_q^{(1)}+
\frac{\Theta_{1,3}}8-\frac{5\Theta_{2,2}}{32}
\end{align}
The associated indicial equation is
\begin{equation}
  \alpha ^3  (\alpha -1) =0,
  \label{indicial-su2squared}
\end{equation}
while the twisted indicial equation
\begin{equation}
    (2 \alpha -1)^3 (2 \alpha +1) = 0
   \label{indicial-su2squared-conjugate} 
\end{equation}
has solutions $\alpha = -1/2, \,1/2, \, 1/2, \,1/2$. Therefore, the leading growth will come from the smallest root $\alpha = -1/2$. Since all four solutions differ by an integer, there could be a priori cubic logarithmic terms. In this case, from the matrix integral analysis we know that we should only obtain quadratic logarithms because the power of the logarithms cannot be larger than the rank of the gauge group.


\section{The $\tau\to-d/c$ expansions}\label{sec:tauToRat}


In this section we define $\tilde{q}$ via
\begin{equation}
    \tilde{q}:=e^{2\pi i\frac{a\tau+b}{c{\tau}+d}}=e^{2\pi i\frac{a}{c}}\,e^{-\frac{2\pi i}{c\tilde{\tau}}},
\end{equation}
where
\begin{equation}
    \widetilde{\tau}:=c\tau+d \qquad \text{ and } \qquad \begin{pmatrix}
        a & b \\ c & d
    \end{pmatrix}\in \mathrm{PSL}(2,\mathbb{Z}),
\end{equation}
which implies that $\gcd(c,d)=1$.

We hope that the integers $a$ and $c$ featuring in the modular transformation $\tau\to\frac{a\tau+b}{c\tau+d}$, subject to $ad-bc=1,$ will not be confused below with the 4d central charges $\mathrm{a}$ and $\mathrm{c}$.

\subsection*{Direct analysis of the matrix integral}


To study the integral \eqref{eq:integralSchur} in the limit $\widetilde{\tau}=c\tau+d\to0$, we use that up to a constant phase $\eta(\frac{a\tau+b}{\tilde{\tau}})=\sqrt{\tilde{\tau}}\eta(\tau),$ to obtain
\begin{equation}
    (q;q)\;= \; \frac{1}{\sqrt{-i \wt{\tau}}}  \;
      e^{ -\frac{2\pi i}{24c}(\wt{\tau}+\frac{1}{\wt{\tau}})}   \, (\tilde{q};\tilde{q}),
\label{eq:PochRationalEst}
\end{equation}
with $\tilde{q}=e^{2\pi i\frac{a\tau+b}{\tilde{\tau}}}=e^{2\pi i\frac{a}{c}}\,e^{-\frac{2\pi i}{c\tilde{\tau}}}$, and up to a constant phase. Then using~\eqref{eq:theta0modularTrans} inside the integrand of~\eqref{eq:integralSchur} we find analogously to \eqref{eq:outExpansion}
the contribution of the outer patch as
\begin{equation}
   \begin{split}
   \mathcal{I}_{\text{out}}(q)
&= e^{-\frac{i\pi \tilde{\tau} }{c}\mathrm{c}} \, e^{-\frac{2\pi i}{c\tilde{\tau}}2(\mathrm{a}-\mathrm{c})}\times\\
&\quad\bigg(\int_{\mathcal{S}'_\epsilon} \frac{\mathrm{d}^{r_G}\boldsymbol{x}}{(-i\tilde{\tau})^{r_G}}\ C_1\  e^{-\frac{2\pi i}{c\tilde{\tau}}\cdot L_c(c\,\boldsymbol{x})}+\cdots+\int_{\mathcal{S}'_\epsilon} \frac{\mathrm{d}^{r_G}\boldsymbol{x}}{(-i\tilde{\tau})^{r_G}}\  C_n\ e^{-\frac{2\pi i}{c\tilde{\tau}}\cdot L_c^{(n)}(c\,\boldsymbol{x})}+\cdots\bigg),
\end{split}\label{eq:outExpansionTilde}
\end{equation}
with $C_n$ constants (\emph{i.e.} independent of $\tilde{\tau},x_j$). The function $L_c(\boldsymbol{x})$ is defined as
\begin{equation}
    L_c(\boldsymbol{x}):=\frac{n^{}_{\rho_0}}{4}\,\vartheta\bigg(\frac{d}{2}\bigg)+\frac{1}{2}\sum_{\rho_+^\chi}\vartheta\left(\rho_+^\chi\cdot\boldsymbol{x}-\frac{d}{2}\right)-\sum_{\alpha_+}\vartheta(\alpha_+\cdot\boldsymbol{x})=:\begin{cases}
   L(\boldsymbol{x})& d\mbox{ even},\\
   L_{t}(\boldsymbol{x})& d\mbox{ odd}.
   \end{cases}\label{eq:tRainsDef}
\end{equation}
We will refer to $L_{t}$ arising for odd $d$ as the \emph{twisted Rains function}.

The functions $L^{(n)}_c$ are given by
\begin{equation} 
L_c^{(n)}(\boldsymbol{x})=L_c(\boldsymbol{x})+\mathbb{Z}_{\ge0}\big(\{\alpha_+\cdot \boldsymbol{x}\},1-\{\alpha_+\cdot \boldsymbol{x}\},\{\rho^\chi_+\cdot \boldsymbol{x}-\frac{d}{2}\},1-\{\rho^\chi_+\cdot \boldsymbol{x}-\frac{d}{2}\},1\big).\label{eq:higherTwistedRains}
\end{equation}
For even $d$ these are the same as the higher Rains functions, and for odd $d$ we will refer to them as the \emph{higher twisted Rains functions}. That is, the $n$th twisted Rains function is the (1st) twisted Rains function plus a linear combination with non-negative integer coefficients of the ``generators'' shown above. Note in particular that $L_c^{(n)}$ is piecewise linear.

The integrals in \eqref{eq:outExpansionTilde} can be evaluated analogously to \eqref{eq:tauTo0Iout} and yield
\begin{equation}
    {q^{\mathrm{c}/2}\,\mathcal{I}_\text{out}(q)=\tilde{q}^{h_{c\,{\text{min}}}} \cdot C_0(\log\tilde{q})^{\mathrm{dim}\mathfrak{h}_{c\,\mathrm{qu}}}+\sum_\ast \tilde{q}^{h_{c\ast}} \cdot C_\ast(\log\tilde{q})^{\mathrm{dim}\mathfrak{h}_c^\ast},}\label{eq:tauToRationalIout}
\end{equation}
where $\tilde{q}=e^{2\pi i\frac{a}{c}}\,e^{-\frac{2\pi i}{c\tilde{\tau}}}$, and $h_{c\,{\text{min}}},h_{c\ast},\mathrm{dim}\mathfrak{h}_{c\,\text{qu}},\mathrm{dim}\mathfrak{h}_c^\ast$ are defined as $h_{\text{min}},h_\ast,\mathrm{dim}\mathfrak{h}_{\text{qu}},\mathrm{dim}\mathfrak{h}^\ast$ if $d$ is even, and defined as before except with $L_t$ instead of $L$ if $d$ is odd. In the latter case we denote them by $h_{t0},h_{t\ast},\mathrm{dim}\mathfrak{h}_{t\,\text{qu}},\mathrm{dim}\mathfrak{h}_t^\ast$.

Although we do not carry out the inner patch analysis in the limit $\tilde{\tau}\to0$, we expect as in the previous section that the expression \eqref{eq:tauToRationalIout} is completed by the inner patch contributions to the form (after including the sum over all $n$)
\begin{align}
q^{\mathrm{c}/2}\,\mathcal{I}(q)&=\tilde{q}^{h_{c\,{\text{min}}}}\cdot P_{\mathrm{dim}\mathfrak{h}_{c\,\text{qu}}}(\log\tilde{q}) +\sum_{n=1}^\infty\sum_{\ast_n} \tilde{q}^{h_{c\ast_n}} \cdot P_{\mathrm{dim}\mathfrak{h}_c^{\ast_n}}(\log\tilde{q})\nn
\\
&=e^{-\frac{2\pi i}{c\tilde{\tau}}[2(\mathrm{a}-\mathrm{c})+L_c^{\mathrm{min}}]}\cdot P_{\mathrm{dim}\mathfrak{h}_{c\,\text{qu}}}(\log\tilde{q}) +\sum_{n=1}^\infty\sum_{\ast_n} e^{-\frac{2\pi i}{c\tilde{\tau}}[2(\mathrm{a}-\mathrm{c})+L_{c\ast}^{(n)}]} \cdot P_{\mathrm{dim}\mathfrak{h}_c^{\ast_n}}(\log\tilde{q}),
\label{eq:tauToRationalFullAllExpSmall}
\end{align}
where $P_{\mathrm{dim}\mathfrak{h}_c^{\ast_n}}(\log\tilde{q})$ is a polynomial of order ${\mathrm{dim}\mathfrak{h}_c^{\ast_n}}$ in $\log\tilde{q}$, and where $\tilde{q}=e^{2\pi i\frac{a}{c}}\,e^{-\frac{2\pi i}{c\tilde{\tau}}}$.\\

\noindent\textbf{Remark.} Note that for even $d$ the singular asymptotic of the Schur index in the new $\tau\to-d/c$ limit is very similar (up to a $\tau\to c\wt{\tau}$ replacement) to the $\tau\to0$ asymptotic of the previous section. But for odd $d$ the twisted Rains function arises which requires separate analysis. We summarize this state of affairs by saying roughly that \emph{there is no new information in the singular asymptotic for even $d$}. On the other hand, all odd $d$ have similar singular asymptotic (up to a $\tau\to c\wt{\tau}$ replacement) to the $d=\pm1,c=1$ case, which is somewhat analogous to the Cardy limit of the $\mathcal{N}=1$ index on the \emph{second sheet} \cite{Cassani:2021fyv}. So we can  say roughly that \emph{all the new information in the singular asymptotic for odd $d$ is contained in the twisted Rains function.} We will see below examples where the twisted Rains function contains no new information either (\emph{i.e.} $L^\mathrm{min}_{t}=L^\mathrm{min}$ and $\mathrm{dim}\mathfrak{h}_{c\,\mathrm{qu}}=\mathrm{dim}\mathfrak{h}_{\mathrm{qu}}$), but also examples where \emph{there is} novel information in the singular asymptotic on the second sheet.\\

\subsection*{MLDE perspective}

As stated previously, the analysis in this case will be different for the untwisted and twisted cases.

When the MLDE is untwisted, the vvmf transforms under the full modular group. Thus, we can map any rational number to $+i\infty$ through a modular transformation to extract the asymptotic growth from the terms with smallest exponents in the vvmf. The growth will be similar to the growth at $\tau\to 0$, and the only changes will be a suppression in the exponent given by the denominator of the rational number $-d/c$, and the mixing of the solutions due to having a different matrix $\rho$. More explicitly, the modular transformation for a vvmf reads
\begin{equation}
  \chi(\tau) = \rho \begin{pmatrix} a & b \\ c & d
    \end{pmatrix}^{-1}
   \chi\left(\frac{a\tau+b}{c\tau+d} \right).
\end{equation}
In the limit $\tilde{\tau}=c\tau+d \to 0$, the leading growth of the vacuum module $\chi^{}_0(\tau)$ will come from the first coefficients of each component of the full vector of solutions, provided that the matrix $\rho$ mixes all components. In general, the asymptotic form of $\chi^{}_0(\tau)$ for $\tau$ near $-d/c$ will be a linear combination given by the matrix $\rho$ of the components $\chi^{}_0,\dots,\chi_{n-1}$ which now will be expanded in $\tilde{q} = e^{2\pi i \frac{a}{c}}e^{-2\pi i \frac{1}{c\tilde{\tau}}}$, and thus the exponent $e^{-2\pi i \frac{1}{c\tilde{\tau}}}$ will be suppressed by $c$ with respect to $e^{-2\pi i \frac{1}{{\tau}}}$ that we had in the limit $\tau \to 0$.

The Schur index associated to a half-integer graded VOA will be a solution to a twisted MLDE. The solutions of this equation in general will not transform into themselves under arbitrary modular transformations, but only under the transformations belonging to the congruence subgroup $\Gamma^0(2)$  (see Appendix \ref{App:appendix-modular} for more details on modular forms for congruence subgroups),
\beq
\G^0(2):=\left\{\left(
\begin{array}{cc}
a&b\\
c&d
\end{array}
\right)\in \mathrm{PSL}(2,\mathbb{Z}),\quad
b= 0\
{\rm mod}\ 2
\right\}.
\label{definition-gamma02}
\eeq

In the $\tau\to 0$ analysis we saw that there is no  transformation belonging to $\Gamma^0(2)$ which maps $0$ to the $+i\infty$ cusp, and therefore we needed to extract the growth from the solutions to the \textit{conjugate} MLDE. 

For an arbitrary rational number $-d/c$, it turns out that sometimes we can find a $\Gamma^0(2)$ element
\begin{equation}
    \left(\begin{array}{cc}
a&b\\
c&d
\end{array}\right), \quad ad-bc=1 \quad \text{ with } \quad b= 0\
{\rm mod}\ 2
\end{equation}
which maps $-d/c$ to $+i\infty$. The unit determinant condition imposes $\gcd(b,d)=1$ and therefore when $d$ is even there will be no such $\Gamma^0(2)$ element and again we need to extract the growth from the solutions of the conjugate MLDE, as in the $\tau \to 0$ case. When $d$ is odd one can always find $b$ which is even\footnote{To see that there is always a $\Gamma^0(2)$ matrix for a given rational $-d/c$ with odd $d$, notice that the following matrix product
\begin{equation}
    \begin{pmatrix}
    1 & 1 \\
    0 & 1
    \end{pmatrix}\begin{pmatrix}
    a & b \\
    c & d
    \end{pmatrix} = \begin{pmatrix}
    a+c & b+d \\
    c & d
    \end{pmatrix}\quad w/\quad b\equiv 1\ {\rm mod\ 2},\ d\in 2\Z+1,\ ad-bc =1
\end{equation}
allows to obtain a matrix in $\Gamma^0(2)$ from a matrix which does not belong to the group (\emph{i.e.} $b\neq 0\ {\rm mod}\ 2$), while leaving the pair $c,d$ fixed.} and therefore the growth is given by the solutions to the MLDE satisfied by the Schur index. Therefore, for $d$ odd the asymptotics will be in general different from the $\tau\to 0$ case.

In summary, the expansion we get for $q$ around a root of unity has the same structure as the expansion \eqref{eq:MLDEstructuralExp} near $q=1,$ except that $\tilde{q}=e^{-2\pi i/\tau}$ in \eqref{eq:MLDEstructuralExp} should be replaced with $\tilde{q}=e^{2\pi i\frac{a}{c}}\,e^{-\frac{2\pi i}{c\tilde{\tau}}}$, and that in the twisted case $\alpha_i$ should be the roots of the conjugate MLDE for odd $d$ but the roots of
the MLDE itself for even $d$.

\subsection*{Examples}

\subsubsection*{The SU($2$) $\mathcal{N}=4$ theory}
\begin{figure}[h!]
\centering
    \includegraphics[scale=.45]{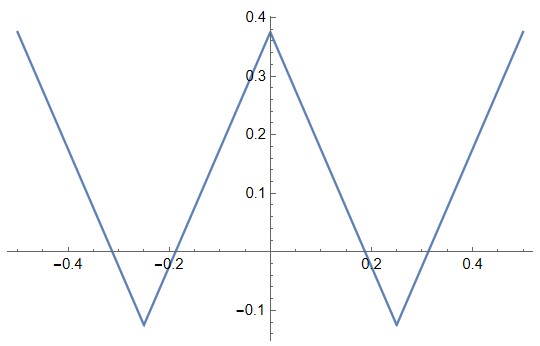}
\caption{Twisted Rains function of SU(2) $\mathcal{N}=4$ SYM.  
\label{fig:n=4tw}}
\end{figure}

The twisted Rains function looks as in Fig~\ref{fig:n=4tw}. So we have $L_{t }^{\mathrm{min}}=-1/8$ and $\mathrm{dim}\mathfrak{h}^{(1)}_{t\,\text{qu}}=0$. Since $\mathrm{c}=\frac{3}{4}$ and $\mathrm{a}-\mathrm{c}=0,$ the $\wt{\tau}\to0$ asymptotic looks like
\begin{equation}
    \mathcal{I}(q)e^{\frac{i\pi\wt{\tau}}{c}\frac{3}{4}}\simeq\begin{cases}
   \frac{A(c,d)}{-i\wt{\tau}}+B(c,d)& d\mbox{ even,}\\
   e^{\frac{2\pi i}{8c\wt{\tau}}}\cdot C(c,d)&d \mbox{ odd,}\end{cases}\label{eq:N=4ratAsy}
\end{equation}
up to exponentially suppressed corrections. The coefficients $A,B,C$ can be obtained using methods of \cite{Ardehali:2021irq}, but we do not compute them here.

In this case the MLDE is invariant only under $\G^0(2)$ transformations and therefore from the previous reasoning the vacuum character will display two types of behaviour near rational points $-d/c$. For $d$ even the growth will depend on the roots of the MLDE satisfied by the vacuum character \eqref{MLDE-su2n4} and for odd $d$ the asymptotic growth will depend on the roots of the conjugate MLDE \eqref{conjugate-MLDE-su2n4}. Since the roots differ by a half-integer and the expansion of the characters is in half-integers, the character associated to the smallest root can contain a logarithmic term, and one can check that this is indeed the case. 

More explicitly, for $\tau$ near $-d/c$
\begin{equation}
    \mathcal{I}(q)e^{\frac{i\pi\wt{\tau}}{c}\frac{3}{4}}\simeq\begin{cases}
   \frac{A(c,d)}{-i\wt{\tau}}+B(c,d)& d\mbox{ even,}\\
   e^{\frac{2 \pi i}{8c\wt{\tau}}}\cdot C(c,d) + e^{-\frac{3}{8}\frac{2 \pi i}{c\wt{\tau}}}\left(\frac{D(c,d)}{-i\tilde{\tau}}+E(c,d) \right) &d \mbox{ odd,}\end{cases}
\end{equation}
consistently with the twisted Rains function analysis.

\subsubsection*{SU($2$) SQCD}
\begin{figure}[h!]
\centering
    \includegraphics[scale=.45]{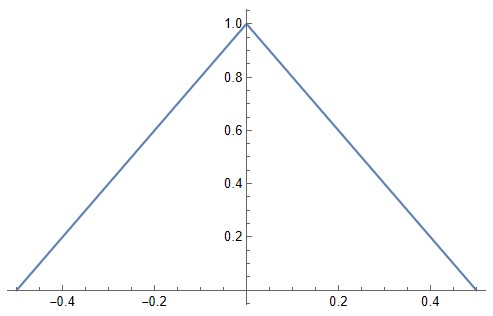}
\caption{Twisted Rains function of SU(2) SQCD.
\label{fig:sqcdtw}}
\end{figure}

The twisted Rains function looks as in Fig~\ref{fig:sqcdtw}. So we have $L_{t }^{\mathrm{min}}=0$ and $\mathrm{dim}\mathfrak{h}^{(1)}_{t\,\text{qu}}=0$. Since $\mathrm{c}=\frac{7}{6}$ and $\mathrm{a}-\mathrm{c}=-\frac{5}{24},$ the small-$\wt{\tau}$ asymptotic looks like
\begin{equation}
    \mathcal{I}(q)e^{\frac{i\pi\wt{\tau}}{c}\frac{7}{6}}\simeq
   e^{\frac{5i\pi}{6c\wt{\tau}}}\cdot C(c,d),
   \label{growth-rains-su2qcd-rationals}
\end{equation}
up to exponentially suppressed corrections.

In this example the MLDE \eqref{su2qcd-MLDE} is untwisted and therefore the analysis is analogous to the $\tau\to 0$ case. More explicitly, for this example the vacuum module can be written as
\begin{equation}
    \chi_1(\tau) = C_1(c,d) \chi_1\left(\frac{a\tau+b}{c\tau+d} \right)+ C_2(c,d) \chi_2\left(\frac{a\tau+b}{c\tau+d} \right),
\end{equation}
where the constants $C_i(c,d)$ come from the matrix $\rho$. Therefore, the small $\tilde{\tau} = c\tau+d$ asymptotics has the form
\begin{equation}
   e^{i\pi{\tau}\frac{7}{6}}\ \mathcal{I}(q)= \chi_1(\tau) = \tilde{C}_1(c,d) (e^{-\frac{7}{12}\frac{2\pi i}{c\tilde{\tau}}}+\dots) + \tilde{C}_2(c,d) (e^{\frac{5}{12}\frac{2\pi i}{c\tilde{\tau}}}+e^{-\frac{7}{12}\frac{2\pi i}{c\tilde{\tau}}}\left(A+\frac{B}{c\tilde{\tau}}\right)+\dots),
\end{equation}
where we have reabsorbed some possible constant factors in $\tilde{C}_i(c,d)$. The growth has the same functional form as for $\tau\to 0$ except for the suppression by a factor of $c$ on the exponents, in accordance with the direct analysis of the matrix integral \eqref{growth-rains-su2qcd-rationals}. Finally, we just mention that from the knowledge of the form of $\rho$ for the $S$ and $T$ transformations given above one can obtain the form of $\rho$ for any $\mathrm{SL}(2,\mathbb{Z})$ matrix from its decomposition in terms of the matrices $S$ and $T$, thus being possible to obtain the exact expressions of all the coefficients for any rational $-d/c$.

\subsubsection*{USp(4) with half-hyper in 16}
\begin{figure}[h]
\centering
    \includegraphics[scale=.45]{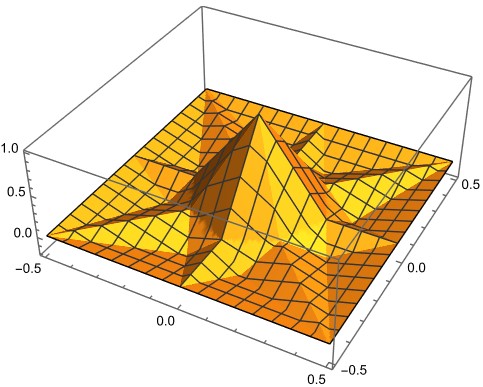}
\caption{Twisted Rains function of the USp(4) theory. 
\label{fig:usp4tw}}
\end{figure}

The twisted Rains function looks as in Fig~\ref{fig:usp4tw}. So we have $L_{t }^{\mathrm{min}}=-1/5$ and $\mathrm{dim}\mathfrak{h}^{(1)}_{t\,\text{qu}}=0$. Since $\mathrm{c}=\frac{7}{3}$ and $\mathrm{a}-\mathrm{c}=\frac{1}{12},$ we have
\begin{equation}
    \mathcal{I}(q)  e^{\frac{i\pi \wt{\tau} }{c}\frac{7}{3}}\simeq e^{\frac{\pi i}{15c\wt{ \tau}}}\cdot C(c,d).
    \label{growth-rains-usp4-rationals}
\end{equation}

For this example the MLDE is invariant under the full modular group and therefore the asymptotic growth near any rational  number $-d/c$ will have the same functional form that for the limit $\tau\to 0$ with an extra $1/c$ suppression in the exponents, as in \eqref{growth-rains-usp4-rationals}.

\subsubsection*{SU$(2)^2$ with one $(2, 2)$, two $(2, 1)$s, and two $(1, 2)$s}
\begin{figure}[h]
\centering
    \includegraphics[scale=.45]{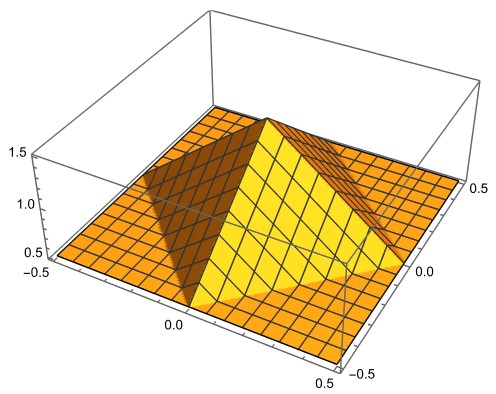}
\caption{Twisted Rains function of the SU(2)$\times$SU(2) theory.
\label{fig:su2xsu2tw}}
\end{figure}

The twisted Rains function looks as in Fig~\ref{fig:su2xsu2tw}. So we have $L_{t }^{\mathrm{min}}=1/2$ and $\mathrm{dim}\mathfrak{h}^{(1)}_{t\,\text{qu}}=1$.
\begin{equation}
    \mathcal{I}(q)e^{\frac{i\pi\wt{\tau}}{c}\mathrm{c}}\simeq\begin{cases} e^{-\frac{2\pi i}{c\wt{ \tau}}(2(\mathrm{a}-\mathrm{c}))}\cdot C(c,d)
   & d\mbox{ even,}\\
   e^{-\frac{2\pi i}{c\wt{ \tau}}(2(\mathrm{a}-\mathrm{c})+\frac{1}{2})}\big(\frac{A(c,d)}{(-i\wt{\tau})^2}+\frac{B(c,d)}{-i\wt{\tau}}+C(c,d)\big) &d \mbox{ odd.}\end{cases}\label{growth-rains-su2xsu2-rationals}
\end{equation}

The MLDE for this example is invariant only under $\G^0(2)$ and therefore we will have two types of asymptotic behaviours. The growth will be determined by the solutions of the indicial equation \eqref{indicial-su2squared}, $\alpha = 0,0,0,1$, and the conjugate indicial equation \eqref{indicial-su2squared-conjugate}, $\alpha = -1/2,1/2,1/2,1/2$. More explicitly, for $\tau\to -d/c$,
\begin{equation}
    \mathcal{I}(q)e^{\frac{i\pi\wt{\tau}}{c}\mathrm{c}}\simeq\begin{cases} e^{\frac{1}{2}\frac{2\pi i}{c\wt{ \tau}}}\cdot C(c,d)
   & d\mbox{ even,}\\
   \frac{A(c,d)}{(-i\wt{\tau})^2}+\frac{B(c,d)}{-i\wt{\tau}}+C(c,d) &d \mbox{ odd,}\end{cases}
\end{equation}
where we have only displayed the leading asymptotic behaviour. The case of odd $d$ displays a quadratic logarithm stemming from the triple degeneracy of the root $\alpha=0$. Note the compatibility with \eqref{growth-rains-su2xsu2-rationals} since $\mathrm{a}-\mathrm{c}=-1/4.$


\section{Discussion}\label{sec:Discussion}

We end with a conceptual summary and some avenues of further exploration.

\subsection{Conceptual summary}\label{subsec:cancelDisc}

We have studied the high-temperature expansion of the Schur index from two perspectives: the matrix-integral (available for Lagrangian theories as in \eqref{eq:indexTheta0}) and the MLDE (available more generally \cite{Beem:2017ooy}). Both approaches allow obtaining general \emph{structural}---or semi-qualitative---results in the sense of determining the structure of the $\tilde{q}$-expansion but not the values of the coefficients. See Eq.~\eqref{eq:tauTo0FullAllExpSmall} obtained from the matrix-integral and Eq.~\eqref{eq:MLDEstructuralExp} obtained from the MLDE perspective, both uniformly convergent over compact subsets of the unit disk $|q|<1.$

We also presented \emph{algorithmic}---or fully quantitative---procedures from both perspectives, for obtaining not only the structure but also the coefficients of the expansion, and exemplified them in deriving Eq.~\eqref{eq:SU(2)N=4RGderivedAsy} for $\mathrm{SU}(2)$ $\mathcal{N}=4$ theory (from the matrix-integral) and Eq.~\eqref{eq:Stransformed-sqcd} for $\mathrm{SU}(2)$ SQCD with $N_f=4$ (from the MLDE).

In the matrix-integral approach, the algorithmic procedure involves decomposing the integration domain to an outer patch where the integrand can be safely Taylor expanded around $\tilde{q}=0$, and various inner patches whose contribution is found by integrating their RG equation, analogously to how the problem of large logarithms is solved in 4d QED (\emph{cf.} Section~7.2.2 of \cite{Burgess:2020tbq}) or $\varphi^4$ theory (\emph{cf.} Section~11.3 of \cite{brezin2010introduction}).

In the MLDE approach, the algorithmic procedure involves finding the multiplier system $\rho:\mathrm{PSL}(2,\mathbb{Z})\to\mathrm{GL}(n,\mathbb{C})$ (of the $n$-dimensional vvmf comprised by the MLDE solutions) on the two generators $T$ and $S$ of the modular group, by examining the power series expressions for the MLDE solutions in the case of $T$ and by demanding that $\rho$ is a homomorphism (\emph{e.g.} using the relations that $S$ and $ST$ satisfy) in the case of $S$, fixing any remaining ambiguity by numerical examination.

Comparing the structural expansions \eqref{eq:tauTo0FullAllExpSmall} and \eqref{eq:MLDEstructuralExp} (as well as their root-of-unity analogs) allows us establish the following two results for Lagrangian $\mathcal{N}=2$ SCFTs:
\begin{itemize}
    \item that the highest power of $\log q$ in the $q$ expansion of any VOA character that mixes with the vacuum character under $\mathrm{PSL}(2,\mathbb{Z})$ is not greater than the rank (alternatively, the degeneracies of the corresponding MLDE indicial roots are not greater than rank$\,+1$);
    \item that the conformal dimension of any VOA character that mixes with the vacuum character under $\mathrm{PSL}(2,\mathbb{Z})$ is a rational number (alternatively, the corresponding MLDE indicial roots are rational).
\end{itemize}
The former follows from the fact that the dimension of the quantum moduli spaces is bounded above by the dimension of the classical moduli space, \emph{i.e.} rank, while the latter follows from Remark~2 below \eqref{eq:tauTo0FullAllExpSmall}, together with the fact that the central charges $\mathrm{a},\mathrm{c}$ of Lagrangian $\mathcal{N}=2$ SCFTs are rational---see \eqref{eq:LagrangianCentralCharges}. It would seem fitting to conjecture that both statements remain true for non-Lagrangian $\mathcal{N}=2$ SCFTs as well\footnote{The latter statement in fact follows very generally from the arguments in \cite{Rastelli:2023sfk} within the MLDE framework, assuming the conjecture that the modular orbit of the vacuum character of the VOA consists only of characters of ordinary simple modules.} (whose rationality of central charges follows from the arguments in \cite{Argyres:2018urp,Caorsi:2018zsq,Martone:2020nsy,Argyres:2020wmq,Rastelli:2023sfk}).

The structural results suffer from the following drawback however: without recourse to more powerful tools it can not be ruled out that some of the coefficients in the structural expansions are actually zero.\footnote{There is one exception: the leading term in \eqref{eq:tauTo0FullAllExpSmall} was proven in \cite{ArabiArdehali:2016fjg} to be nonzero.} While in any \emph{special case} we can run the algorithmic procedures to determine whether various coefficients are zero or not, \emph{in general} the best we can do is to combine the structural results with naturalness arguments that would indicate vanishing of certain coefficients as unnatural. This is how we argued from \eqref{eq:tauTo0FullAllExpSmall} and \eqref{hstarOrig} that the term
\begin{equation}
    \tilde{q}^{2(\mathrm{a}-\mathrm{c})},\label{eq:a-cTerm}
\end{equation}
should naturally be present in the high-temperature expansion of the Schur index of all Lagrangian theories, even if it does not correspond to the leading asymptotics. (It therefore seems fitting to conjecture presence of \eqref{eq:a-cTerm} in the $\tilde q$ expansion of the Schur index of all $\mathcal{N}=2$ SCFTs.)

Let us now discuss a case where the structural result \eqref{eq:tauTo0FullAllExpSmall} predicts certain powers of $\tilde{q}$, but due to \emph{natural cancellations} the corresponding coefficients end up vanishing, so actually those powers are absent.

$\mathrm{SU}(2)$ SQCD with $N_f=4$ provides such an example. The error term in \eqref{eq:SQCDtauTo0ExpSmallFirstSub} following from the higher Rains function $L^{(3)}(x)=L(x)+\{x\}$ as in Figure~\ref{fig:su2sqcdHigher3}, ends up cancelling against similar contributions arising from other higher Rains functions. Specifically, from $L^{(4)}(x):=L(x)+(1-\{x\})$ as well as
\begin{equation}
\begin{split}
    &L^{(3)}(x)+\{2x\},\quad L^{(3)}(x)+2\{2x\},\quad L^{(3)}(x)+(1-\{2x\}),\quad L^{(3)}(x)+2(1-\{2x\}),\\
    & L^{(4)}(x)+\{2x\},\quad L^{(4)}(x)+2\{2x\},\quad L^{(4)}(x)+(1-\{2x\}),\quad L^{(4)}(x)+2(1-\{2x\}).
\end{split}\label{eq:higherRainsListSQCD}
\end{equation}
Let us check the cancellation on the outer patch. There we can Taylor expand \eqref{eq:indexStransformed} and compare with \eqref{eq:outExpansion} to read $C_3=-4$ and $C_4=-4$, as well as
\begin{equation}
    \begin{split}
     &+8,\qquad -4,\qquad +8,\qquad-4,\\
     &+8,\qquad-4,\qquad+8,\qquad-4,
    \end{split}
\end{equation}
for the other higher Rains functions listed in \eqref{eq:higherRainsListSQCD}. The integrals inside the parenthesis in \eqref{eq:outExpansion} can then be easily evaluated, for instance with Mathematica, and the would-be error terms (corresponding to $\tilde{q}^{3/2}$ inside the parenthesis) are seen to cancel.

We consider these cancellations natural because the $q$- (or low-temperature) expansion of the Schur index in this case contains only integer powers, indicating the MLDE should be untwisted, implying that its other solutions $\chi^{}_j$ must have integer power series also---up to the overall $\tilde{q}^{h_j}$ of course. In other words, ``modular symmetry'' considerations in this case rule out terms of the form $\tilde{q}^{\alpha+\text{odd}/2}$, with $\alpha$ either the vacuum character root $\alpha_0=\mathrm{c}/2$ (naturally present from the MLDE perspective), or the root $\alpha_1$ (naturally present from the matrix-integral perspective) that ought to be less than $2(\mathrm{a}-\mathrm{c})$ by an integer. This argument would fail if the MLDE had another indicial root $\alpha_2$ differing from $\alpha_0$ or $\alpha_1$ by half an odd integer of course; in the $\mathrm{SU}(2)$ SQCD case we know from explicit examination that $\alpha_{0,1}$ exhaust the indicial roots.

It would be desirable to understand more systematically the set of all symmetries of the index that could underlie natural cancellations in its high-temperature expansion. For analogous recent attempts to find symmetry explanations of various cancellations (``magic zeros'') in perturbative particle physics see \emph{e.g.}~\cite{Craig:2021ksw}.

\subsection{Criteria for validity of the $\mathrm{a}-\mathrm{c}$ formula}

The leading piece of the expansion \eqref{eq:tauTo0FullAllExpSmall} corresponds to $\tilde{q}^{h_0}$. In many cases it turns out that $L^{\mathrm{min}}=0$, thus $h_0=2(\mathrm{a}-\mathrm{c})$. The leading asymptotic of the Schur index then becomes
\begin{equation}
    \mathcal{I}(q)\approx e^{-\frac{2\pi i}{ \tau}(2(\mathrm{a}-\mathrm{c}))}.\label{eq:BNasymptotic}
\end{equation}
This is the asymptotic encountered for certain Schur indices first by Buican and Nishinaka in \cite{Buican:2015ina}, and is analogous to the Di~Pietro-Komargodski (DK) formula for $\mathcal{N}=1$ indices \cite{DiPietro:2014bca}. The latter formula is observed to be valid in many examples where $\mathrm{a}-\mathrm{c}<0$, which is most often the case, but violated in a few cases where $\mathrm{a}-\mathrm{c}>0$ \cite{ArabiArdehali:2015ybk,DiPietro:2016ond,Hwang:2018riu}. We also saw in Section~\ref{sec:tauTo0} the $\mathrm{USp}(4)$ example with $\mathrm{a}-\mathrm{c}>0$ where \eqref{eq:BNasymptotic} is violated.\footnote{ For Lagrangian $\mathcal{N}=2$ SCFTs the DK formula is valid iff \eqref{eq:BNasymptotic} is valid \cite{ArabiArdehali:2015ybk}.} In other words
\begin{equation}
   \text{the $\mathrm{a}-\mathrm{c}$ asymptotics are violated}\stackrel{\text{in all ex.}}{\Longrightarrow} \text{$\mathrm{a}>\mathrm{c}$}.\label{statement:c-aAndDK}
\end{equation}
The converse is not true: puncture-less $\mathrm{SU}(2)$ class-$\mathcal{S}$ theories of genus $g\ge2$ have $\mathrm{a}>\mathrm{c}$ but do satisfy \eqref{eq:BNasymptotic}, see \cite{ArabiArdehali:2016fjg}. It would be interesting however to find criteria under which the converse does apply.

One generally true statement regarding validity of \eqref{eq:BNasymptotic} (or the $\mathcal{N}=1$ DK formula) is as follows. Assume that the dimensionally reduced theory is ``strictly bad'' (\emph{cf}.~\cite{Gaiotto:2008ak}), namely that with the naive assignment of $R$-charges as descending from four dimensions there are BPS monopole operators of strictly negative $R$-charge.\footnote{If the 3d theory does not have monopoles of strictly negative $R$-charge, but has monopoles of zero $R$-charge, we refer to it as ``marginally bad'', to distinguish it from the \emph{strictly bad} cases.} Since the function $\tilde{L}(\boldsymbol{x})$ in Eq.~\eqref{eq:Ltilde} (or its $\mathcal{N}=1$ analog in \cite{ArabiArdehali:2015ybk}) encodes the naive $R$-charges of BPS monopoles (see \emph{e.g.} \cite{Kapustin:2010mh}), it will be negative in the direction corresponding to the monopole with strictly negative $R$-charge. This implies that the Rains function $L(\boldsymbol{x})$ will be negative in that direction as well. Hence $L^{\text{min}}<0$, and the formula \eqref{eq:BNasymptotic} is violated. In other words
\begin{equation}
  \text{the dimensionally reduced theory is strictly bad}\Longrightarrow \text{the $\mathrm{a}-\mathrm{c}$ asymptotics are violated}.\label{statement:c-aAndBadReduc}
\end{equation}
Whether the converse is true in general or not is an open problem.

Combined with the empirical observation that in all examples where the $\mathrm{a}-\mathrm{c}$ asymptotics are violated we have $\mathrm{a}>\mathrm{c},$ one is led to
\begin{equation}
  \text{the dimensionally reduced theory is strictly bad}\stackrel{\text{in all ex.}}{\Longrightarrow} \text{$\mathrm{a}>\mathrm{c}$}.\label{statement:conjBadC-A}
\end{equation}
It would be interesting to investigate more systematically this relation and the criteria under which its converse is true. Potential relevance of the intriguing recent work \cite{Shan:2023xtw} to establishing such criteria is currently being investigated.\\

Incidentally, while the $\mathcal{N}=1$ index $\mathcal{I}^{}_{\mathcal{N}=1}(p,q)$ has been observed to have a universal asymptotic on the ``second sheet'' (\emph{i.e.} as $p\to1,\, q\to e^{\pm2\pi i}$) governed by $3\mathrm{c}-2\mathrm{a}$ \cite{Cassani:2021fyv,Cabo-Bizet:2019osg,Kim:2019yrz}, the Schur index does not exhibit similar universality on its second sheet. This can be seen by noting that already the four examples in \eqref{eq:N=4ratAsy}, \eqref{growth-rains-su2qcd-rationals}, \eqref{growth-rains-usp4-rationals}, and \eqref{growth-rains-su2xsu2-rationals} with $c=1,\,d=\pm1$, do not accommodate a universal formula of the form $\mathcal{I}\approx e^{\frac{f(\mathrm{a},\mathrm{c})}{\tilde\tau}}$ for any linear function $f$: just put $f(\mathrm{a},\mathrm{c})=\gamma_1\mathrm{c}+\gamma_2 \mathrm{a}+\gamma_3$, find the $\gamma$s with Mathematica from the first three of our examples, and observe that the fourth example does not satisfy $\mathcal{I}\approx e^{\frac{f(\mathrm{a},\mathrm{c})}{\tilde\tau}}$ with the $\gamma$s so obtained. To put it more analogously to the $\mathcal{N}=1$ story, the robust asymptotic on the second sheet of the Schur index is really $\mathcal{I}\approx e^{\frac{0}{\tilde\tau^2}+\mathcal{O}(\frac{1}{\tilde\tau})}$, with $0$ being the universal coefficient.

\subsection{Field theory explanation of the exponentially small corrections}

The leading exponential and its dressing polynomial have been given EFT explanation in~\cite{Ardehali:2021irq}, building on the pioneering work of Di~Pietro and Komargodski \cite{DiPietro:2014bca} and its subsequent developments in \cite{DiPietro:2016ond,Cassani:2021fyv,ArabiArdehali:2021nsx}. It would be nice to obtain an EFT explanation of the  exponentially suppressed corrections.

The leading exponential has been related to the vacuum energy in a crossed channel; see Appendix~B of \cite{ArabiArdehali:2019zac} and Section~6.3 of \cite{Dorey:2023jfw}, both leveraging Shaghoulian's ``modularity''~\cite{Shaghoulian:2016gol}. It should be possible to interpret the exponentially suppressed terms in the expansion as arising from massive BPS particles in the crossed channel.\footnote{We are grateful to Z.~Komargodski for suggesting this possibility to us.} This is also presently under~study.

\begin{acknowledgments}

We are indebted to A.~Deb, M.~Dedushenko, G.~Eleftheriou, Z.~Komargodski, M.~Litvinov, R.~Yazdi, S.~Murthy, L.~Rastelli, B.~Rayhaun, H.~Rosengren, G.~Zafrir for helpful discussions. We are especially grateful to S.~Razamat for initial collaboration and sharing with us the conjecture~\eqref{eq:ShlomoConj}, as well as M.~Dedushenko, Z.~Komargodski, and L.~Rastelli for instructive comments on a draft of this paper. MR thanks the ICISE in Quy
Nhon, Vietnam for hospitality and is grateful to the organisers of the Pollica Summer Workshop supported by the Regione Campania, Università degli Studi di Salerno, Università degli Studi di Napoli ``Federico II", the Physics Department ``Ettore Pancini" and ``E.R. Caianiello", and Istituto Nazionale di Fisica Nucleare. AA was supported in part by the NSF grant PHY-2210533 and the Simons Foundation grants 397411 (Simons Collaboration on the Nonperturbative Bootstrap) and 681267 (Simons Investigator Award). MR acknowledges financial support from the grants ERC Consolidator Grant N. 681908, ``Quantum black holes: A macroscopic window into the microstructure of gravity'' and the Royal society grant RF/ERE/210168 which is part of the Royal Society URF grant ``The Atoms of a de Sitter Universe''. The work of MM is supported by STFC grant ST/T000759/1.\\



\end{acknowledgments}

\appendix

\section{Special functions}\label{app:special}

The elliptic gamma function is defined as
\begin{equation}
    \Gamma(z;p,q):=\prod_{j,k\ge 0}\frac{1-z^{-1}p^{j+1}q^{k+1}}{1-z
    p^{j}q^{k}}.\label{eq:GammaDef}
\end{equation}
We denote it for $p=q$ by $\Gamma(z),$ keeping the dependence on $q$ implicit.

The theta function is defined via
\begin{equation}
    \theta_0(u;\tau)=\prod_{k=0}^\infty (1-e^{2\pi i(u+k\tau)})(1-e^{2\pi i(-u+(k+1)\tau)}).
\end{equation}

The following identities are helpful in simplifications
\begin{equation}
    \Gamma(qz)=\theta_0(z;q)\Gamma(z),\qquad \theta_0(q^{1/2}z;q)=\theta_0(q^{1/2}z^{-1};q),\qquad \Gamma(z)\Gamma(z^{-1})=\frac{1}{\theta_0(z;q)\theta_0(z^{-1};q)}.\label{eq:simpGamThet}
\end{equation}

The modular property
\begin{equation}
    \theta_0(u;\tau)=e^{-\frac{i\pi}{\tau}(u^2+u+\frac{1}{6})+i\pi(u+\frac{1}{2})-\frac{i\pi\tau}{6}}\theta_0(u/\tau;-1/\tau),
\end{equation}
implies (by replacing $u$ with $\{u\}_\tau$ on the RHS) the product formula
\begin{equation}
    \theta_0(u;\tau)=-ie^{-\frac{i\pi}{6}(\tau+\frac{1}{\tau})}e^{\frac{i\pi}{\tau}\{u\}^{}_\tau(1-\{u\}^{}_\tau)}e^{i\pi\{u\}^{}_\tau}\times\prod_{k=1}^\infty (1-e^{-\frac{2\pi i}{\tau}(k-\{u\}^{}_\tau)})(1-e^{-\frac{2\pi i}{\tau}(k-(1-\{u\}^{}_\tau))}).\label{eq:theta0S}
\end{equation}
Here $\{u\}^{}_\tau:=u-\lfloor \mathrm{Re}u-\cot(\arg\tau)\mathrm{Im}u\rfloor$. Replacement of $u$ with $\{u\}_\tau$ is justified because $\theta_0(u,\tau)$ is 1-periodic in $u$.

For our purposes two cases are of special interest. They are
\begin{equation}
    \theta_0(\tau/2;\tau)=2\,e^{\frac{i\pi\tau}{12}-\frac{i\pi}{6\tau}}\prod_{k=1}^\infty (1+\tilde{q}^k)^2,\label{eq:theta0t/2S}
\end{equation}
as well as \eqref{eq:theta0S} with $u=x+r\tau$, and $x$ real and away from integers, with $r\in\{0,1/2\}$, so that we can replace \begin{equation}
\{u\}^{}_\tau\to\{x\}+r\tau,    
\end{equation}
where $\{x\}=x-\lfloor x\rfloor;$ then we get
\begin{equation}
    \theta_0(x+\frac{\tau}{2};\tau)=e^{-\frac{i\pi}{6}(-\frac{\tau}{2}+\frac{1}{\tau})}e^{\frac{i\pi}{\tau}\vartheta(x)}\times\prod_{k=1}^\infty (1+e^{-\frac{2\pi i}{\tau}(k-\{x\})})(1+e^{-\frac{2\pi i}{\tau}(k-(1-\{x\}))}),\label{eq:theta0shiftedS}
\end{equation}
where $\vartheta(x):=\{x\}(1-\{x\}).$

The $q$-Pochhammer symbol is defined as
\begin{equation}
    (z;q):=\prod_{k=0}^{\infty}(1-zq^k).\label{eq:PochDef}
\end{equation}

To obtain the most general modular transformation of $\theta_0$, we start with that of $\theta_1$, defined via
\begin{equation}
    \theta_1(u;\tau)=iq^{1/8}e^{-i\pi u}(q;q)\theta_0(u;\tau).\label{eq:theta0theta1}
\end{equation}
The general transformation formula for $\theta_1$ reads (up to a constant phase)
\begin{equation}
   \theta_1(\frac{u}{\tilde{\tau}};\frac{a\tau+b}{c\tau+d})=\sqrt{-i\tilde{\tau}}e^{i\pi c u^2/\tilde{\tau}}\theta_1(u;\tau),
\end{equation}
where $\tilde{\tau}=c\tau+d$. This, using the fact that up to a constant phase $\eta(\frac{a\tau+b}{\tilde{\tau}})=\sqrt{\tilde{\tau}}\eta(\tau),$ implies
\begin{equation}
\begin{split}
\theta_1(u;\tau)&=e^{-i\pi c u^2/\tilde{\tau}}e^{\frac{2\pi i}{8}\frac{a\tau+b}{\tilde{\tau}}}e^{-i\pi u/\tilde{\tau}}e^{-\frac{2\pi i}{24}\frac{a\tau+b}{\tilde{\tau}}}q^{1/24}(q;q)\,\theta_0(\frac{u}{\tilde{\tau}};\frac{a\tau+b}{c\tau+d})\\
\Longrightarrow \theta_0(u;\tau)&=e^{-i\pi c u^2/\tilde{\tau}-i\pi u/\tilde{\tau}+i\pi u}\,e^{-\frac{1}{12}2\pi i\frac{1}{c\tilde{\tau}}}e^{-\frac{1}{12}2\pi i\frac{\tilde{\tau}}{c}}\,\theta_0(\frac{u}{\tilde{\tau}};\frac{a\tau+b}{c\tau+d})\\
&=e^{-\frac{i\pi}{6c}(\tilde{\tau}+\frac{1}{\tilde{\tau}})}\, e^{\frac{i\pi}{c\tilde{\tau}}\{cu\}_{\tilde{\tau}}(1-\{cu\}_{\tilde{\tau}})+i\pi u}\big(1-e^{-\frac{2\pi i}{c\tilde{\tau}}\{cu\}_{\tilde{\tau}}}\big)\prod_{k=1}^\infty \big(1-e^{2\pi i k\frac{a}{c}}\,e^{-\frac{2\pi i}{c\tilde{\tau}}(k\pm\{cu\}_{\tilde{\tau}} )}\big),\label{eq:theta0modularTrans}
\end{split}
\end{equation}
again up to a constant phase.

\section{MLDEs and logarithmic vector-valued modular forms} \label{App:appendix-modular}

\subsection*{Modular forms}

Here we collect a variety of facts, definitions and conventions on modular forms and modular differential operators, for more details see \emph{e.g.} \cite[Appendix A]{Beem:2017ooy} or one of the standard references \cite{ranestad20081}.

The modular group $\G\equiv \mathrm{PSL}(2,\Z)$ is defined as the subgroup of matrices $\gamma$ in $\mathrm{SL}(2,\mathbb{R})$ with entries restricted to take integer values:
\begin{equation}
	\gamma = \begin{pmatrix}
		a & b \\ c & d
	\end{pmatrix},\qquad a, b, c, d\in \Z,\qquad ad-bc=1
\end{equation}
 and where each matrix $\gamma$ is identified with $-\gamma$. This group acts on the modular parameter $\tau\in\H$ taking values in the upper half plane
\beq
\gamma:\t\to \gamma\circ\tau = \frac{a\t+b}{c\t+d}.
\eeq
The modular group is generated by two elements:
\beq
S:\t\to-\frac1\t,\qquad T:\t\to \t+1.
\eeq
We define the \emph{nome} as $q:=e^{2\pi i\t}$ and the modular group acts on it, $\gamma\circ q$, by its action on $\tau$. 

The principal congruence subgroups of $\G$ are defined as:
\beq
\G(N):=\left\{\left(
\begin{array}{cc}
a&b\\
c&d
\end{array}
\right)\in \G,
\ a= d=\pm1,\
b= c= 0\
{\rm mod}\ N
\right\}
\eeq
while congruence subgroups are subgroups of $\G$ which contain $\G(N)$ as a subgroup. We are particularly interested in the following congruence subgroup:
\beq
\G^0(2):=\left\{\left(
\begin{array}{cc}
a&b\\
c&d
\end{array}
\right)\in \G,\quad
b= 0\
{\rm mod}\ 2
\right\},
\label{gamma02}
\eeq
as well as its $S$-conjugate subgroup $\G_0(2) = S\, \G^0(2)\,S$.

A modular form of weight $k\in\mathbb{Z}$ for $\G$ is a holomorphic function $f:\H\to \C$ which transforms as
\beq
f\left(\frac{a\t+b}{c\t+d}\right)=(c\t+d)^kf(\t),\qquad \left(
\begin{array}{cc}
a&b\\
c&d
\end{array}
\right)\in \G,
\eeq
and is finite in the limit ${\rm Im}\,\t\to +\infty$. It follows from the invariance under $T$ transformations that any modular form has a convergent Fourier expansion in $q$ and is finite in the $q\to0$ limit:
\beq
f(\t)=\sum_{n=0}^\infty a_nq^n.
\eeq
Relaxing the growth condition at infinity allowing the modular form to contain a finite number of terms with negative $q$ exponents in the Fourier expansion defines a weakly holomorphic modular form. The space of modular forms of weight $k$ is denoted by $M_k(\G)$ and it is a standard result that the dimension of $M_k(\G)$ is finite for any $k$, non-zero for even $k>2$, while for $k=0$ it only contains constant functions.

Similarly we can define modular forms for $\tilde{\G}=\G_0(2),\G^0(2)\subset\G$ as a holomorphic function on $\H$ for which 
\beq
f\left(\frac{a\t+b}{c\t+d}\right)=(c\t+d)^kf(\t),\qquad \left(
\begin{array}{cc}
a&b\\
c&d
\end{array}
\right)\in \tilde{\G},
\eeq
and which has subexponential growth at the cusps of $\tilde{\G}$ (\emph{i.e.}, at all orbits of $\mathbb{P}^{1}(\mathbb{Q})$ under $\tilde{\G}$). The previous Fourier expansion remains unchanged for modular forms in $\G_0(2)$. For the group $\G^0(2)$, any modular form  has a Fourier series expansion in $q^{1/2}$ instead
\beq\label{twiCon}
f(\t)=\sum_{n=0}^{\infty}a_n q^{n/2}.
\eeq

The ring of modular forms for the full modular group $\G$ is freely generated by $\E_4(\t)$ and $\E_6(\t)$:
\beq
\bigoplus_{k=0}^{\infty}M_k(\G)=\C[\E_4(\t),\E_6(\t)],
\eeq
where the ordinary Eisenstein series are modular forms for the full modular group $\G$ of weight $2k$, with $k\geq 2$, defined as
\beq
\E_{2k}(\t):=-\frac{B_{2k}}{2k!}+\frac{2}{(2k-1)!}\sum_{n\geq1}\frac{n^{2k-1}q^n}{1-q^n},
\eeq
where $B_{2k}$ is the 2$k$-th Bernoulli number. The case of $\E_2(\t)$ is special as $\E_2(\t)$ is not a modular form but a quasi-modular form since it transforms anomalously under modular transformations:
\beq
\E_2\left(\frac{a\t+b}{c\t+d}\right)=(c\t+d)^2\E_2(\t)-\frac{c (c\t+d)}{2\pi i}.
\label{eq:e2anomalous}
\eeq

For the case of $\G_0(2)$ and $\G^0(2)$, the following results hold:
\begin{align}
\bigoplus_{k=0}^\infty M_k(\G_0(2)) &=\C[\theta_2(\t)^4,\theta_3(\t)^4]^{\cS_2} \\
\bigoplus_{k=0}^\infty M_k(\G^0(2)) &=\C[\theta_3(\t)^4,\theta_4(\t)^4]^{\cS_2}
\end{align}
where $\theta_i(\t)$ are the classical Jacobi theta functions defined as:
\begin{align}
\vartheta_{00}(\t)&\equiv\theta_3(\t):=\sum_{n=-\infty}^\infty q^{\frac{n^2}2}\\
\vartheta_{01}(\t)&\equiv\theta_4(\t):=\sum_{n=-\infty}^\infty (-1)^nq^{\frac{n^2}2}\\
\vartheta_{10}(\t)&\equiv\theta_2(\t):=\sum_{n=-\infty}^\infty q^{\frac12\left(n+\frac12\right)^2}
\end{align}
and the superscript $\cS_2$ means that we are restricting to symmetrized polynomials. We can therefore define the following basis:
\begin{align}
\Theta_{r,s}(\t)&:=\theta_2(\t)^{4r}\theta_3(\t)^{4s}+\theta_2(\t)^{4s}\theta_3(\t)^{4r},\quad r\leq s\\
\widetilde{\Theta}_{r,s}(\t)&:=(-1)^{r+s}\left(\theta_2(\t)^{4r}\theta_3(\t)^{4s}+\theta_2(\t)^{4s}\theta_3(\t)^{4r}\right),\quad r\leq s
\end{align}
in terms of which we have
\begin{align}
M_{2k}(\G^0(2))&={\rm span}\{\Theta_{r,s}(\t)|r+s=k\}\\
M_{2k}(\G_0(2))&={\rm span}\{\widetilde{\Theta}_{r,s}(\t)|r+s=k\}.
\end{align}
Even though the $S$-transformation does not belong to $\G^0(2)$ or $\G_0(2)$, its action on the previous basis elements is known, it maps elements of one basis to the other:
 \begin{equation}
     \Theta_{r,s}\left(-\frac{1}{\tau}\right) = \tau^{2r+2s} \widetilde{\Theta}_{r,s}(\tau).
     \label{eq:s-transformed-thetas}
 \end{equation}

The homomorphism
\begin{equation}
	\rho: \Gamma \to \mathrm{GL}(n,\mathbb{C}),
\end{equation}
denotes an $n$-dimensional representation of $\Gamma$. A vector-valued modular form of weight $k\in\mathbb{Z}$ and multiplier system $\rho$ is a function $\chi = (\chi_0,\dots,\chi_{n-1}): \mathbb{H}\to \mathbb{C}^n $ which transforms as
\begin{equation}
    \chi(\gamma\tau) = (c\tau+d)^k\rho(\gamma)\chi(\tau), \hspace{4mm} \gamma \in \Gamma,
    \label{eq:defn-vvmf}
\end{equation}
and its component functions are bounded as $\rm{Im}\tau \to \infty$. If the component functions have exponential growth in this limit then $\chi(\tau)$ is referred to as a weakly holomorphic vector-valued modular form.

Similarly, the vector-valued modular forms for the congruence subgroups $\G^0(2)$, $\G_0(2)$ with a multiplier system $\rho$ are defined by demanding that they transform in the same way, \eqref{eq:defn-vvmf}, as their counterparts for the full modular group but now restricting it to $\G^0(2)$ or $\G_0(2)$. Moreover, depending on the growth of the component functions they are holomorphic or weakly holomorphic vvmfs.

A weakly holomorphic logarithmic vector-valued modular form of weight $k\in\Z$ with multiplier system $\rho$ is a holomorphic function $\chi = (\chi_0,\dots,\chi_{n-1}): \mathbb{H}\to \mathbb{C}^n $ satisfying the transformation property \eqref{eq:defn-vvmf} and such that its $q$ expansion contains logarithms of $q$.


\subsection*{Modular Linear Differential Equations}

The $\tau$-derivative of a modular form does not transform covariantly under modular transformations. More explicitly, one can check that for $f\in M_{k}(\G)$ the $\tau$ derivative transforms in the following way:
\beq
q\frac d {dq}f(\t)\ \to \ (c\t+d)^{k+2}q\frac d {dq}f(\t)+k \frac{c(c\t+d)^{k+1}}{2\pi i}f(\t),
\eeq
where $q\frac d {dq} = \frac{1}{2\pi i}\frac{\partial}{\partial\tau} $. 

The anomalous transformation of the quasi-modular form $\E_2$ can be used to define a differential operator which maps modular forms of weight $k$ to modular forms of weight $k+2$:
\beq
\partial^{(k)}:=q\frac d{dq}+k\E_2(\t).
\eeq
We are now in a position to define a differential operator of degree $k$ which acts on weight 0 modular forms:
\beq
D_q^{(k)}:=\partial^{(2k-2)}\circ\partial^{(2k)}\circ...\circ\partial^{(0)},
\eeq
and maps them to modular forms of weight $2k$. The operator $D_q^{(0)}$ is the identity operator by convention. 

Since these differential operators transform with weight $2k$ under the action of $\G$ or any congruence subgroup
\begin{equation}
    D_{\gamma\circ q}^{(k)} = (c\tau+d)^{2k}D_q^{(k)}, \qquad \gamma = \begin{pmatrix}
        a & b \\ c & d
    \end{pmatrix}\in {\G},
\end{equation}
we can construct a large class of modular linear differential operators of a fixed weight $2k$ as sums of operators of weight $2k-2r$ multiplied by modular forms of weight $2r$. We are particularly interested in those that are holomorphic and monic
\beq\label{diffop}
\cD^{(k)}_{\tilde{\G}}\equiv D_q^{(k)}+\sum_{r=1}^kf_r(\tau)D_q^{(k-r)},\quad f_r(\tau)\in M_{2r}(\tilde{\G}),
\eeq
where $\tilde{\G}\in\{\G,\G^0(2),\G_0(2)\}$. Monic means the coefficient of $D^{(k)}$ is one. By holomorphic we mean that the modular forms $f_r(\tau)$ defining the coefficients of the MLDO belong to $M_{2r}(\tilde{\G})$ and therefore have finite value at $q\to 0$. In the following we will refer to the cases $\tilde{\G}=\G$ and $\tilde{\G}=\G_0(2),\G^0(2)$ as the \emph{untwisted} and \emph{twisted} case, respectively. The elements of the kernel of an MLDO define a modular linear differential equation (MLDE):
\begin{equation}
    \cD^{(k)}_{\tilde{\G}}f(\tau) =\left( D_q^{(k)}+\sum_{r=1}^kf_r(\tau)D_q^{(k-r)}\right)f(\tau) = 0.
\end{equation}
From the modular properties of the differential operators, one can show that $k$ linearly independent solutions to an MLDE define a rank $k$ vector-valued modular form \cite{mason-vvmf-mlde}.

Finally, consider an MLDE for the congruence subgroup $\Gamma^0(2)$:
\beq\label{diffop-gamma-naught}
\left(D_q^{(k)}+\sum_{r=1}^kf_r(\tau)D_q^{(k-r)}\right)f(\tau) = 0,\quad f_r(\tau)\in M_{2r}(\G^0(2)),
\eeq
where each $f_r(\tau)$ is expressed as a linear combination of functions $\Theta_{s,t}(\t)$ with $s+t=r$. Due to the transformation \eqref{eq:s-transformed-thetas}, the $S$-transformed MLDE, called the conjugate MLDE, is given by changing $\Theta_{s,t}(\t)$ for $\widetilde{\Theta}_{s,t}(\t)$, and keeping the same coefficients in the linear combinations. Writing the modular forms defining the new MLDE as $\tilde{f}_r(\tau)\in M_{2r}(\G_0(2))$, the conjugate equation reads
\beq\label{diffop-gamma-naught-conjugate}
\left(D_q^{(k)}+\sum_{r=1}^k\tilde{f}_r(\tau)D_q^{(k-r)}\right)f(\tau) = 0,\quad \tilde{f}_r(\tau)\in M_{2r}(\G_0(2)).
\eeq

\subsection*{Vvmfs and MLDEs}


As explained in the text, the (appropriately normalized) Schur index of any four-dimensional $\mathcal{N} = 2$ SCFT is conjectured to solve a finite-order, monic, holomorphic, (twisted) modular linear differential equation \cite{Beem:2017ooy}. In particular, this implies that the Schur index is a component of a vector-valued modular form.

To obtain the full vector one has to find the full set of solutions of the MLDE. Such MLDEs are easily seen (for instance from \eqref{eq:MLDEdiffExpanded} below and its analogs in the twisted case) to have only a regular singularity at $q=0$ inside the unit disk \cite{mason-vvmf-mlde}, and therefore they are Fuchsian ordinary differential equations \cite{ince1949ordinary}. The complete set of solutions to such equations can be obtained using the Frobenius method by implementing the recursion relations starting from the solutions of the indicial equation. When the roots of the indicial polynomial are not equal mod $ \mathbb{Z}$, the solutions will be given in terms of power series. When two roots differ by an integer, some solutions will consist of sums of power series and power series times logarithms of the MLDE variable.

In the following, we explain the steps more explicitly for the case of the modular group. We are interested in the set of functions which are mapped to zero by operators of the form \eqref{diffop},
\beq
\cD^{(k)}_{{\G}}\equiv D_q^{(k)}+\sum_{r=1}^kf_r(q)D_q^{(k-r)},\quad f_r(q)\in M_{2r}({\G}).
\eeq 
The operators $D_q^{(j)}$ can be written as
\begin{equation}
	D_q^{(j)} = \sum_{i=0}^k Q_{2i,j} q^{j-i} \frac{d^{j-i}}{dq^{j-i}},
\end{equation}
where $ Q_{2i,j} $ are polynomials in the Eisenstein series $\E_2,\E_4$ and $\E_6$ with $Q_{0,j}=1$, and therefore the full MLDO can be written as 
\begin{equation}
	\cD^{(k)}_{{\G}}=  \sum_{i=0}^k R_{2i} q^{k-i} \frac{d^{k-i}}{dq^{k-i}}.\label{eq:MLDEdiffExpanded}
\end{equation}
The point $q=0$ is a regular singular point since $R_{2i}(\tau)$ are again polynomials of Eisenstein series and therefore they are regular at $i\infty$ (see \cite{mason-vvmf-mlde} for more details). To solve these Fuchsian equations the standard procedure is to first obtain the indicial polynomial associated to it and find its roots. More concretely, the power series ansatz
\begin{equation}
	\cD^{(k)}_{{\G}}\left( q^\alpha \sum_{n=0}^\infty a_n q^n\right)=0,\quad \alpha \in \R,
\end{equation}
when $q=0$ is a regular singularity can be rewritten as
\begin{equation}
	q^{\alpha-k}(\alpha(\alpha-1)\cdots(\alpha-k+1)a_0 + \dots) + q^{\alpha-k+1}(\dots) + \dots=0,
\end{equation}
and the coefficient of $q^{\alpha-k}$ is the indicial polynomial. For a Fuchsian equation, the degree of the indicial polynomial is equal to the order of the equation, $k$. If the roots of the polynomial $\alpha_1, \dots, \alpha_k$ are all different and do not differ by an integer, the set of series
\begin{equation}
	q^{\alpha_i}\phi_i(q)=q^{\alpha_i}\sum_{n=0}^\infty a_{i,n} q^n,
\end{equation}
is the full set of linearly independent solutions, where the coefficients $a_{i,n}$ are fixed by the recursion relations given by the MLDE.

If some roots of the indicial polynomial differ by an integer, the previous set of power series might not be linearly independent. In this case, it can be shown that the inclusion of logarithms and powers of logarithms multiplying the power series with smaller roots is enough to get the full set of linearly independent solutions. 

More precisely, consider the example of a second order MLDE with roots $\alpha_1,\alpha_2$ that differ by an integer $\alpha_1 = \alpha_2 + n$, $n\in \mathbb{N}_0$. One can show (see, \emph{e.g.} \cite{ince1949ordinary}) that 
\begin{equation}
	q^{\alpha_1} \phi_1(q), \quad \phi_1(q) = \sum_{i\geq 0} a_i q^i,
\end{equation}
with $a_0\neq 0$ is a solution to the MLDE and that the other solution has the form
\begin{equation}
	q^{\alpha_2} \phi_2(q) + A\log(q) q^{\alpha_1} \phi_1(q), \quad \phi_2(q) = \sum_{i\geq 0} b_i q^i,
\end{equation}
with $b_0 \neq 0$. For $\alpha_1 = \alpha_2$, the constant $A$ is necessarily non-zero. For $\alpha_1 > \alpha_2$, the constant $A$ can be zero depending on the recursion relations derived from the MLDE. A relevant point to notice for our analysis is that the logarithms appear multiplying the term with the largest root.

In general, the solution of a monic holomorphic MLDE of degree $n$ is an $n$-dimensional vector with components of the form
\begin{equation}
	\chi_i(\tau) = q^{\alpha_i}\sum_{j=0}^{n-1}\log^j(q)\phi_{i,j}(q),
\end{equation}
where $\phi_{i,j}(q)$ are power series, and therefore the general vector of solutions $\chi(\tau)$ is a logarithmic vector-valued modular form. When all $\phi_{i,j}(q)$ with $j>0$ are zero, $\chi(\tau)$ is a holomorphic vector-valued modular form.

The repeated roots of the indicial equation get translated in the multiplier system associated to the vector-valued modular form by making $\rho(T)$ not diagonalizable. This can be seen from the fact that the transformation $\tau \to \tau+1$ mixes the solutions associated to the same root due to the logarithmic terms in the vector components.

\bibliographystyle{JHEP}
\bibliography{references}

\end{document}